\documentclass{article}
\pdfoutput=1
\usepackage{jcappub}
\usepackage{graphicx}
\usepackage{color}
\usepackage{multirow}
\usepackage{amsmath}
\usepackage{amssymb}
\usepackage{epstopdf}

\def\lbldef#1#2{\expandafter\gdef\csname #1\endcsname {#2}}

\def\href#1#2{#2}

\title{Constraints on deviations from $\Lambda$CDM within Horndeski gravity}
\author[a]{Emilio Bellini,}
\author[a]{Antonio J. Cuesta,}
\author[b,a,c,d]{Raul Jimenez,}
\author[b,a,c,e]{Licia Verde}

\affiliation[a]{ICCUB, University of Barcelona (IEEC-UB), Mart{\'\i} i Franqu{\`e}s 1, E08028 Barcelona, Spain}
\affiliation[b]{ICREA (Instituci\'o catalana de recerca i estudis avan\c{c}ats)}
\affiliation[c]{Radcliffe Institute for Advanced Study, Harvard University, MA 02138, USA}
\affiliation[d]{Institute for Applied Computational Science, Harvard University, MA 02138, USA}
\affiliation[e]{Institute of Theoretical Astrophysics, University of Oslo, 0315 Oslo, Norway}

\abstract{Recent anomalies found in cosmological datasets such as the low multipoles of the Cosmic Microwave Background or the low redshift amplitude and growth of clustering measured by e.g., abundance of galaxy clusters and redshift space distortions in galaxy surveys, have motivated explorations of models beyond standard $\Lambda$CDM. Of particular interest are models where general relativity (GR) is modified on large cosmological scales. Here we consider deviations from $\Lambda$CDM+GR within the context of Horndeski gravity, which is the most general theory of gravity with second derivatives in the equations of motion. We adopt a parametrization in which the four additional Horndeski functions of time $\alpha_i(t)$ are proportional to the cosmological density of dark energy $\Omega_{DE}(t)$. Constraints on this extended parameter space using a suite of state-of-the art cosmological observations are presented for the first time. Although the theory is able to accommodate the low multipoles of the Cosmic Microwave Background and the low  amplitude of fluctuations 
from redshift space distortions, we find no significant tension with $\Lambda$CDM+GR when performing a {\it global} fit to recent cosmological data and thus there is no  evidence against $\Lambda$CDM+GR from an analysis of the value of the Bayesian evidence ratio of the modified gravity models with respect to $\Lambda$CDM, despite introducing extra parameters.
The posterior distribution of these extra parameters that we derive  return strong constraints on any possible deviations from $\Lambda$CDM+GR in the context of Horndeski gravity. 
We illustrate how  our results
can be applied to a more general frameworks of modified gravity models.}

\begin{document}

\maketitle
\hypersetup{pageanchor=true}

\section{Introduction}
\label{sec:intro}

The latest observations of the cosmic microwave background (CMB) by the Planck satellite \cite{Planck2013,Planck2013b, Planck2015a, Planck2015b} and of large structure clustering by several surveys e.g., \cite{Anderson:2013zyy, Blake2011} have provided us with a better understanding of the Universe. Indeed, despite the large increase in data volume and accuracy, the $\Lambda$CDM model seems to be extremely successful at describing all these data. The $\Lambda$CDM model is based on six free parameters that are adjusted to the observations, and assumes homogeneity, isotropy and that the law of gravity is Einstein's general relativity (GR). However, as data have improved, some
``inconsistencies", ``tensions", ``anomalies",  (renamed ``curiosities" in \cite{Planck2013})  with the $\Lambda$CDM model have appeared and while they seem to be not statistical significant, they are ``persistent". These ``anomalies" are: {\it i)} the low $\ell$ multipoles in the observed CMB  temperature anisotropy power spectrum are lower than predicted by the standard model,  {\it ii)} CMB (Planck) direct measurements of the lensing potential are  mildly in tension with the prediction of the  angular power spectrum best-fitting  $\Lambda$CDM model ,  {\it iii)} the redshift space distortion of  galaxy clustering data, the BOSS Lyman$-\alpha$ power spectrum and the {\it iv)} CFHTLens constraint on the present-day amplitude of fluctuations at linear scales of 8 Mpc/h, $\sigma_8$, are also lower than predicted by the CMB-calibrated  $\Lambda$CDM model.  {\it v)} The observed cluster abundance seems also to be lower than predicted by the model, which could also be explained by a lower $\sigma_8$. However the level of tension with $\Lambda$CDM depends somewhat on how the cluster sample is selected, with Planck SZ-selected clusters giving the stronger tension \cite{PlanckSZclusters} . Finally {\it vi)} the value of the Hubble constant inferred from the CMB within a  $\Lambda$CDM model is lower than the value directly measured locally \cite{Riess2011} and e.g., Ref.~\cite{tension} and refs therein.\footnote{The $\sigma_8$ tension  is what drives claims in the literature for massive neutrinos e.g., \cite{BOSSneutrino, PlanckSZclusters} or phantom dark energy \cite{PlanckDEpaper}, the $H_0$ tension drives  claims for sterile neutrinos \cite{Hamann:2010bk}.}

 These anomalies are  each at the $\approx 2\sigma$ level and therefore taken individually are not of enough statistical significance as to claim any ``new physics'' beyond the standard $\Lambda$CDM model. 
However, if all these anomalies could be fit by a single model, they would become more than ``curiosities"
and the possible discovery window for new physics opens up. Indeed, there has been already significant activity in the literature to try to explain these anomalies \footnote{A search the ADS database reports in excess of 100 articles addressing this issue.}.

One basic assumption of the $\Lambda$CDM model is its reliance on Einstein's general relativity theory. One obvious step is then to investigate if changing this assumption could provide a better fit to the current observational data. The models proposed in the literature that deviate from Einstein's gravity are abundant, and in principle infinite. Then, in the absence of strong theoretical priors that can identify a particular model as the ``real'' theory of gravity, a generic framework is highly preferred. However, increasing the generality of this framework increases also its complexity. It is then crucial to balance our theoretical beliefs in order to optimize the freedom that is left to fit the observations. To achieve this description many attempts have been done in the literature, such as the parametrized post-Friedmann \cite{Tegmark:2001zc} or an Effective Field Theory  approach \cite{Gubitosi:2012hu, Bloomfield:2012ff}. In the present paper we assume that modifications of gravity on cosmological scales are well described by an additional scalar degree of freedom with at most second-order derivatives in the covariant equations of motion. An additional requirement is that the theory should satisfy the weak equivalence principle, i.e.\ all matter species are coupled minimally and universally to the metric $g_{\mu\nu}$. This leads to the Horndeski lagrangian \cite{Horndeski:1974wa, Deffayet:2009mn, Kobayashi:2011nu}. It encompasses many of the classical  dark energy (DE) and Modified Gravity (MG) models studied to explain the late-time cosmic acceleration: quintessence, kinetic gravity braiding, galileons, $f (R)$.  \footnote{Note that with the Horndeski lagrangian it is possible to describe particular classes of models that break the weak equivalence principle, such as the ones studied in~\cite{Gleyzes:2015pma,Gleyzes:2015rua}. However, general frameworks as non-universal disformal couplings, or Lorentz-invariance-violating models (e.g., Horava-Lifshitz gravity), or beyond Horndeski theories, e.g.,~\cite{Zumalacarregui:2012us,Koivisto:2012za,Zumalacarregui:2013pma,Gleyzes:2014dya,Gao:2014fra} are not described by the Horndeski lagrangian. These models therefore are not considered here.}

An efficient description of the linear perturbation theory in the Horndeski lagrangian has been investigated in two independent ways in \cite{Gubitosi:2012hu, Bloomfield:2012ff, Gleyzes:2013ooa, Gleyzes:2014qga, Gleyzes:2014rba} and in \cite{BS} (see also \cite{Bellini:2015wfa} for an equivalent approach at second-order). We will make use of the notation of the latter, but the two results are equivalent. The maximal freedom one can have is described by four functions of time besides the Hubble parameter $H(t)$, which is responsible for the expansion history of the universe, and one constant, i.e.,~the amount of matter density today. One of the advantages of this approach is that we can separate the background evolution from the perturbations. Indeed, these four functions appear only at the perturbative level, they are independent from each other and with respect to $H(t)$. Here we choose a particular parametrization  for these functions proposed by \cite{BS} and used in \cite{Bellini:2015wfa}, where  they are proportional to the cosmological density of dark energy $\Omega_{DE}(t)$. This reduces the effective freedom of the theory to four free parameters. 

The aim of this paper is to constrain the free parameters of our description using the most recent CMB and large scale structure data.

Horndeski gravity includes GR as a special case, thus if  Horndeski gravity provides a significantly  better statistical fit to the data  away from the GR limit this will signal possible  new physics. Conversely if not such signal is found the results can be used to place limits of possible deviations from GR on cosmological scales. 
 
Within the $\Lambda$CDM model and its popular extensions it is quite widespread to perform global fits  considering compilations of state-of-the-art cosmological datasets. Specific models, such as $f\left(R\right)$ or galileons, have been considered or specific data sets e.g.~\cite{Kunz:2015oqa,Soergel:2014sna, Hu:2015rva,Raveri:2014cka,Dossett:2014oia,Munshi:2014tua,Hu:2014sea,Dossett:2015nda,Barreira:2014jha,Barreira:2013jma}. This however has not been attempted  systematically for generic deviations from GR such as Horndeski models. Recently, the Planck collaboration \cite{PlanckMG} studied the implications of some DE/MG models with the latest Planck data. In particular, they first parametrized the background evolution only, letting the perturbations evolve with standard equations on top of a non-$\Lambda$CDM background. Then, they parametrized directly the perturbations introducing two independent functions of the curvature perturbations \cite{Simpson:2012ra}, using the public available code \textsc{MGCamb} \cite{Zhao:2008bn,Hojjati:2011ix}. Besides the analysis of some ``real'' theory of gravity, such as massive gravity, $f\left(R\right)$ or coupled quintessence, they also studied DE/MG with the EFT  approach described in the previous paragraph. They used the public code \textsc{EFTCamb} \cite{Hu:2013twa,Raveri:2014cka}, to get constraints on a sub-class of the Horndeski class of models, i.e.~in our notation the alphas different from zero are $\alpha_\mathrm{K}$ and $\alpha_\mathrm{B}=-\alpha_\mathrm{M}$ (see \cite{BS} and Sec.~\ref{sec:method} for details). Their conclusion is that there are no significant deviations from $\Lambda$CDM for all the models investigated. This can be ascribed to the lack of more precise/accurate data, or to the theoretical priors they are imposing, i.e.~not enough freedom in the parameter space. While for the first limit we have to wait for the next generation of surveys, the latter can be relaxed to cover more generic frameworks. This is the scope of this work. In addition here  for the first time we present a  global fit for Horndeski gravity  considering  a compilation of state-of-the-art cosmological data including the latest large-scale structure data.
 
The rest of this paper is organized as follows. In \S \ref{sec:method} we review the Horndeski parametrization and outline the methodology adopted and the data sets used. Results   and their interpretation are reported in \S \ref{sec:results}. Finally we conclude  in \S \ref{sec:conclu}.
\section{Methodology}
\label{sec:method}
We begin by reviewing  the  Horndeski model and the role of the free functions of the model. Since  with cosmological data it is not possible to constrain non-parametrically these functions we parametrize them  and motivate our choice. We also present the data-sets we use here. We consider CMB data and  large-scale structure data obtained from galaxy surveys. In this paper we do not consider clusters data,  weak lensing  data or local measurements of the  Hubble   parameter.
The amount of  tension between the  $\Lambda$CDM model  and cluster abundance depends on the  cluster data set chosen and more specifically on the mass-observable relation, indicating that  this  probe is not as mature and robust  as other  probes of $\sigma_8$. We therefore leave the cluster abundance for future work.
Traditional --2D-- weak lensing analysis uses  non-linear scales  which could be affected by poorly understood physical processes (e.g., baryonic effects). Restricting the analysis to linear scales increases significantly the error-bars making all ``tension"  disappear \cite{Kilbinger:2012qz,Heymans:2013fya, Kitching3dLensing}.  
Moreover the  minimal  description of the Horndeski class of models  as it has been  studied in the literature is strictly valid for linear and second-order perturbations \cite{Bellini:2015wfa}. Even if the goal of modeling non-linear scales  in  generic theories beyond GR is in principle achievable, i.e.,~it is possible to find a finite set of alphas at every order in perturbation theory, it is not developed enough and therefore using non-linear scales is well beyond the scope of this paper.
Finally the tension with the direct $H_0$ measurement  turns out not to be too significant after  recent reanalysis see e.g., \cite{Cuestaladder} and refs therein for discussion.

\subsection{The Horndeski model and adopted parametrization}

The Horndeski action is the most general action for a single scalar field that has second-order equations of motion on any background and satisfies the weak equivalence principle.
By using an Effective Field Theory approach, which has been developed for Horndeski models in \cite{Jimenez:2011nn, Gubitosi:2012hu, Bloomfield:2012ff,Gleyzes:2014dya,Gleyzes:2014qga,Gleyzes:2014rba,Gleyzes:2013ooa,BS} it is possible to identify the minimum number of operators that fully specify the linear evolution of cosmological perturbations.  The result  \cite{BS} is that the maximal freedom of the Horndeski lagrangian at linear-order in perturbation theory can be described efficiently by five functions of time plus one constant, i.e.,~the amount of matter density today $\Omega_{m}(z=0)$. One of the functions is the Hubble parameter $H(t)$, which is responsible for the expansion history of the universe. The other four functions we use appear only at the perturbative level, and they were introduced first in \cite{BS}. In general Horndeski theories $\Omega_{m}(z=0)$ can be computed only through measurements of the large scale structure \cite{Amendola:2012ky,Motta:2013cwa}. This is due to the fact that at the background level DE/MG can have a component that mimics some matter species. As an example, there can be a model that predicts that the DE/MG density scales as $\rho_{DE}(t)=\rho_{DE0}+\rho_{DE1} a(t)^{-3}$, where $\rho_{DE0}$ and $\rho_{DE1}$ are two free parameters. Then, there would be no clear separation between matter and DE just observing the expansion history of the universe, and fixing $H(t)$ would not be enough to determine the matter content today. However, this degeneracy could be removed at the perturbative level by using large scale structure measurements. In this paper, for simplicity we assume that both $H(t)$ and the matter content today are specified by a fiducial $\Lambda$CDM model, i.e.,~the time dependence of $\Omega_{DE}(z)$ has no components $\propto\Omega_m(z)$.

Here, it is useful to summarize the physical properties of these functions, in order to understand how to parametrize them and to interpret the results of this paper:
\begin{itemize}
 \item kineticity: $\alpha_\mathrm{K}$. It is the most standard kinetic term present in perfect fluid structures (it is the only function different from zero in simple DE models as quintessence). It is possible to demonstrate that, if the quasi-static (QS) limit is valid, it does not enter the equations of motion. This indicates that it can not be constrained for models with speed of sound close to the speed of light \cite{SB-quasistaticpaper}. Indeed, in this case the sound horizon of the theory coincides with the cosmological horizon and the QS approximation would be validated on the scales of interest.
 \item braiding: $\alpha_\mathrm{B}$.\footnote{As studied in \cite{Bettoni:2015wta}, the braiding in General Horndeski theories has interesting features at the non-linear level.} It comes from a mixing between the kinetic terms of the metric and the scalar. The more classical models that have a braiding different from zero are $f(R)$/Brans-Dicke and cubic galileons. It modifies the growth of perturbations, i.e.~$f$ and $G_{\rm eff}$. As shown in \cite{BS}, it is responsible for all the new scale dependence of the theory beyond the usual Jeans length. If $\alpha_\mathrm{B}=0$, it is still possible to have anisotropic stress between the curvature perturbations, but $G_{\rm eff}\simeq 1$. For this reason, this property is crucial in order to modify the shape of the power spectrum, and the low $\ell$ multipoles in the observed CMB spectrum.
 \item Effective Planck-mass and Planck-mass run rate: $M_*^2$, $\alpha_\mathrm{M}$. A fixed redefinition of the Planck mass can not have a detectable effect. Indeed, it can be hidden in the field equations by appropriately redefining the densities.  For this reason in this work we will not pay too much attention to $M_*^2$. What we can observe is a variation of the Planck-mass, i.e.\ the Planck-mass run-rate. An evidence of $\alpha_\mathrm{M}\neq0$, indicates that the theory of gravity is non-minimally coupled. As shown in \cite{Saltas:2014dha}, $\alpha_\mathrm{M}$ generates anisotropic stress in the curvature perturbations and modifies the evolution of the gravitational waves. In the presence of non-zero braiding it contributes to modify the growth of perturbations through a modification in the effective Newton's constant.
 \item tensor speed excess: $\alpha_\mathrm{T}$. It is defined as the excess in the speed of tensors with respect to the speed of light. Even if this is its main property, which in principle could be observed \cite{Moore:2001bv,Elliott:2005va,Kimura:2011qn}, on the scalar sector it introduces anisotropic stress and it modifies the growth of perturbations if $\alpha_\mathrm{B}\neq 0$.
\end{itemize}

Hence, $\alpha_\mathrm{B}$ is responsible for all the new scale dependence of the theory beyond the Jeans length. If $\alpha_\mathrm{B}=0$, the effective Newton's constant, which regulates the relation between the curvature perturbation (the trace of the spatial metric) and the matter overdensities, assumes the standard value on sub-horizon scales ($k^2\gg a^2 H^2$), see Eqs.\ (4.9) and (4.11) of \cite{BS}. Then, the evolution of linear perturbations on sub-horizon scales in the extreme QS limit (which is a usual approximation assumed when semi-analytical calculations are performed) can be considered as the evolution of the perturbations in standard perfect fluid structures, with the possibility to have a non-zero anisotropic stress in the curvature perturbations. As noticed in \cite{BS}, it is then possible to define the braiding scale $k_B(t)$ as
\begin{equation}
\frac{k^2_B(t)}{a(t)^2H(t)^2} = \frac{9}{2}\Omega_m(t)+2\left(\frac{3}{2}+\frac{\alpha_K(t)}{\alpha^2_B(t)}\right)\left(\alpha_M(t)-\alpha_T(t)\right)\,.\label{braidingscale}
\end{equation}

Note that the formulation of the braiding scale here is different from the formula given in~\cite{BS}, since we are assuming a $\Lambda$CDM background. The ratio between $\alpha_\mathrm{K}$ and $\alpha^2_\mathrm{B}$ regulates the magnitude of $k^2_B(t)$. In the absence of the second term of Eq.\ (\ref{braidingscale}), the braiding scale lies at the cosmological horizon. But in the presence of non-minimal coupling (i.e.\ $\alpha_\textrm{M}\neq\alpha_\textrm{T}\neq 0$), it is sufficient to tune $\alpha^2_\mathrm{B}\ll \alpha_\mathrm{K}$ in order to push the braiding scales to very small scales (smaller than the scales we are observing). At this point one could ask what the magnitude of $k^2_B(t)$ is in a pure $\Lambda$CDM universe. Indeed, if we consider a DE model like  quintessence ($\alpha_\mathrm{B}=0$), as the closest model to $\Lambda$CDM, then $k^2_B(t)\rightarrow\infty$. This would indicate that standard gravity is recovered on super-braiding scales, where the equations of motion have a perfect fluid structure. On the contrary, we could reduce the  Horndeski class of models to a simple MG theory, such as $f(R)$ ($\alpha_\mathrm{K}=0$), and consider it as the closest model to $\Lambda$CDM. In the limit where the scalar field associated acquires a large mass it would be decoupled from gravity ($f(R)\rightarrow R$), leading to standard results. In this example we could conclude that standard gravity is recovered on sub-braiding scales. Then, the correct answer to the previous question is simply that the braiding scale defines the transition between two different physics. The exact $\Lambda$CDM limit has an indeterminate braiding scale.

When considering standard results in the literature, some confusion arises because, in general, non-minimally coupled models (i.e., $\alpha_\textrm{M}\neq 0$ or $\alpha_\textrm{T}\neq 0$) have also a non-zero braiding. Then, it becomes natural to associate $\alpha_\textrm{K}$ with DE, and $\alpha_\textrm{B}$ with MG. However, in our framework, all these operators are independent  from each other, which ultimately means that MG can lie equivalently on super-braiding or on sub-braiding scales. We conclude that it is easier to have standard gravity on super-braiding scales because it is sufficient to choose $\alpha_\textrm{M}=\alpha_\textrm{T}=0$. On sub-braiding scales the Planck-mass run-rate and the tensor speed excess have to be tuned to suppress the effect of the braiding on the growth of perturbations.

The description we use has five functions of time. One of them, i.e.\ $H(t)$, fully specifies the expansion history of the universe, while the other four modify the evolution of the linear perturbations. 

However, current data are not sufficient to constrain non-parametrically all these functions of time. Therefore we  assume $\Lambda$CDM as our fiducial model for the evolution of the background. At the perturbative level for deviations from  $\Lambda$CDM, we follow the parametrization proposed in \cite{BS,Bellini:2015wfa}, which reads
\begin{equation}
 \alpha_i\left(t\right)\equiv \Omega_{DE}\left(t\right) c_i\,.
\end{equation}
Here, all the $\alpha$ functions are fixed to be proportional to the dark energy density thus reducing the extra freedom of the theory to four parameters, i.e.\ $\{c_\textrm{K},c_\textrm{B},c_\textrm{M},c_\textrm{T}\}$. With this choice we ensure the standard evolution of the universe during the eras dominated by radiation and matter, while we allow for deviations from standard gravity at late-times. This assumption, even if it is not unique,  is motivated as follows:
The $\alpha$-functions  defined in terms of the Horndeski functions (see Appendix A of \cite{BS}) are driven by the same functions of the scalar field as the energy density and its derivatives. Thus, one would naively expect the parameters $c_i$ to be $\mathcal{O}(1)$. In addition, this parametrization describes exactly the behavior of simple shift-symmetric models as the imperfect fluid on its attractor \cite{Pujolas:2011he}, and approximately more complicated models as the best fit of covariant galileons \cite{Barreira:2014jha,Bellini:2015oua}.

In Table \ref{tab:lcdm} we list the various limits of the Horndeski class of models varying the alpha functions. In addition, we give a prediction for the magnitude of the braiding scale when it can be estimated. The $\Lambda$CDM limit is recovered exactly by setting all $c_i=0$. However, $c_\textrm{K}$ represents the contribution given by perfect fluid structures, i.e.,\ what it is usually classified as DE. Then, even if we set $c_\textrm{B}=c_\textrm{M}=c_\textrm{T}=0$ and leave $c_\textrm{K}$ free to vary, the theory of gravity would still be GR. 
\begin{table}
\begin{center}
\begin{tabular}{|c|c|c|c|c|}
\hline 
$\alpha_{\textrm{K}}$ & $\alpha_{\textrm{B}}$ & $\alpha_{\textrm{M}}$ \& $\alpha_{\textrm{T}}$ & $(k_{\textrm{B}}/aH)^{2}$ & Gravity type ($k^{2}\gg a^{2}H^{2}$)\tabularnewline
\hline 
\hline 
\multirow{4}{24pt}{$\rightarrow 0$} & \multirow{2}{24pt}{$\rightarrow 0$} & $\rightarrow 0$ & Indeterminate & Quasi-$\Lambda\textrm{CDM}$ (GR)\tabularnewline
\cline{3-5} 
 &  & $\mathcal{O}\left(1\right)$ & $\rightarrow\infty$ & ``Strong'' MG (non-GR)\tabularnewline
\cline{2-5} 
 & \multirow{2}{24pt}{$\mathcal{O}\left(1\right)$} & $\rightarrow 0$ & $\sim\mathcal{O}\left(1\right)$ & Minimal MG (non-GR)\tabularnewline
\cline{3-5} 
 &  & $\mathcal{O}\left(1\right)$ & $\sim\mathcal{O}\left(1\right)$ & Classical MG (non-GR)\tabularnewline
\hline 
\multirow{4}{24pt}{$\mathcal{O}\left(1\right)$} & \multirow{2}{24pt}{$\rightarrow0$} & $\rightarrow0$ & $\rightarrow\infty$ & Classical DE (GR)\tabularnewline
\cline{3-5} 
 &  & $\mathcal{O}\left(1\right)$ & $\rightarrow\infty$ & Generalized DE (non-GR)\tabularnewline
\cline{2-5} 
 & \multirow{2}{24pt}{$\mathcal{O}\left(1\right)$} & $\rightarrow0$ & $\sim\mathcal{O}\left(1\right)$ & Minimal MG (non-GR)\tabularnewline
\cline{3-5} 
 &  & $\mathcal{O}\left(1\right)$ & $\sim\mathcal{O}\left(1\right)$ & Horndeski (non-GR)\tabularnewline
\hline 
\multicolumn{1}{c}{} & \multicolumn{1}{c}{} & \multicolumn{1}{c}{} & \multicolumn{1}{c}{} & \multicolumn{1}{c}{}\tabularnewline
\hline 
\multicolumn{2}{|c|}{$\alpha_{\textrm{K}}/\alpha_{\textrm{B}}$} & $\left(\alpha_{\textrm{M}}-\alpha_{\textrm{T}}\right)/\alpha_{\textrm{B}}$ & $(k_{\textrm{B}}/aH)^{2}$ & Gravity type ($k^{2}\gg a^{2}H^{2}$)\tabularnewline
\hline 
\hline 
\multicolumn{2}{|c|}{} & $\rightarrow0$ & $\sim\mathcal{O}\left(1\right)$ & Minimal MG\tabularnewline
\cline{3-5} 
\multicolumn{2}{|c|}{$\rightarrow0$} & $\mathcal{O}\left(1\right)$ & $\sim\mathcal{O}\left(1\right)$ & Classical MG\tabularnewline
\cline{3-5} 
\multicolumn{2}{|c|}{} & $\rightarrow\infty$ & Indeterminate & ``Strong'' MG\tabularnewline
\hline 
\multicolumn{2}{|c|}{} & $\rightarrow0$ & Indeterminate & Quasi-$\Lambda\textrm{CDM}$\tabularnewline
\cline{3-5} 
\multicolumn{2}{|c|}{$\rightarrow\infty$} & $\mathcal{O}\left(1\right)$ & $\rightarrow\infty$ & Classical DE\tabularnewline
\cline{3-5} 
\multicolumn{2}{|c|}{} & $\rightarrow\infty$ & $\rightarrow\infty$ & Generalized DE\tabularnewline
\hline 
\multicolumn{1}{c}{} & \multicolumn{1}{c}{} & \multicolumn{1}{c}{} & \multicolumn{1}{c}{} & \multicolumn{1}{c}{}\tabularnewline
\cline{3-5} 
\multicolumn{2}{c|}{} & \multicolumn{2}{c|}{Parameters} & Gravity type\tabularnewline
\hline 
\multicolumn{2}{|c|}{$k^{2}\gg k_{\textrm{B}}^{2}$} & \multicolumn{2}{c|}{$\{\alpha_{\textrm{B}},\alpha_{\textrm{M}},\alpha_{\textrm{T}}\}$} & MG\tabularnewline
\hline 
\multicolumn{2}{|c|}{$k^{2}\ll k_{\textrm{B}}^{2}$} & \multicolumn{2}{c|}{$\{\alpha_{\textrm{K}},\alpha_{\textrm{M}},\alpha_{\textrm{T}}\}$} & Generalized DE\tabularnewline
\hline 
\end{tabular}
\caption{We show the general behavior of the Horndeski class of models varying the functions $\alpha_i$ or a combination of them. In the top table we assume each $\alpha_i$ to be ``$\rightarrow 0$'' or ``$\mathcal{O}\left(1\right)$'', we compute the estimated magnitude of $k_\mathrm{B}$ and we give the expected type of gravity in the deep sub-horizon limit. In the central table we show the magnitude of $k_\mathrm{B}$ and the expected gravity type as a function of particular combination of the $\alpha_i$ functions; this particular combination is closely related to the definition of the braiding scale, Eq.~(\ref{braidingscale}). In the bottom table we give the general behavior and the list of the important parameters on sub/super-braiding scales. This clearly indicates that we expect two different physics in these two regimes.\label{tab:lcdm}}
\end{center}
\end{table}

\subsection{Analysis method}

We use a Bayesian approach to infer the constraints on our model given the datasets  described in Sec.~\ref{sec:data}. The standard fitting of models defined in a multi-dimensional parameter space is done through Monte Carlo sampling techniques. These allow us to infer the posterior  of  the model's parameters, which we obtain by sampling the likelihood of the parameters values given the data and assuming a uniform prior in those parameters.

In order to do this, we use the code \textsc{HiClass}\footnote{\url{http://hiclass-code.net}} \cite{hi-class:2015} to solve the Boltzmann equations. This code, used for the first time in \cite{Bellini:2015oua}, is an extension of the Cosmic Linear Anisotropy Solver Software (CLASS)\footnote{\url{http://class-code.net}} by \cite{Lesgourgues:2011re} to include Horndeski class of modified gravity theories. Since as outlined above we choose the parametrization  in which the time evolution of the four free functions of this class of theories is given by the evolution of $\Omega_{DE}(z)$, the constraints quoted in this paper are actually on the \textit{coefficients} that fix the proportionality between $\alpha$ and $\Omega_{DE}$.

This Boltzmann code is then interfaced by the cosmology sampler \textsc{MontePython} \footnote{\url{http://montepython.net/}} by \cite{Audren:2012wb} which is a powerful engine to run Monte Carlo Markov  chains (MCMC). We run this code using the Metropolis-Hastings sampler to obtain our chains until we consider them converged. In this paper our convergence criterion is given by a value of $R-1\simeq 0.03$ of the Gelman and Rubin parameter \cite{GelmanRubin}. In order to avoid instabilities, we also assume a hard prior on the coefficient $c_\textrm{T}>-0.9$\footnote{For values  $c_\textrm{T}<-0.9$ instabilities in the scalar sector prevent our MCMC to explore this region of parameter space. This region is anyway disfavoured because observations of binary pulsars place a strong limit on the lower bound of the speed of tensors, i.e.\ $c_\textrm{T}\gtrsim-10^{-2}$ \cite{Perenon:2015sla,Jimenez:2015bwa}.} and a initial Planck mass of $M^2_*\geq 1$. We also assume in our runs two massless and one massive neutrino with $m_{\nu}=0.06$eV. Finally, we remove a small fraction of the points which correspond to a negative braiding scale squared $k^2_B(z=0)$.

\subsection{Datasets}
\label{sec:data}
\begin{figure}
\begin{center}
\includegraphics[width=0.45\textwidth]{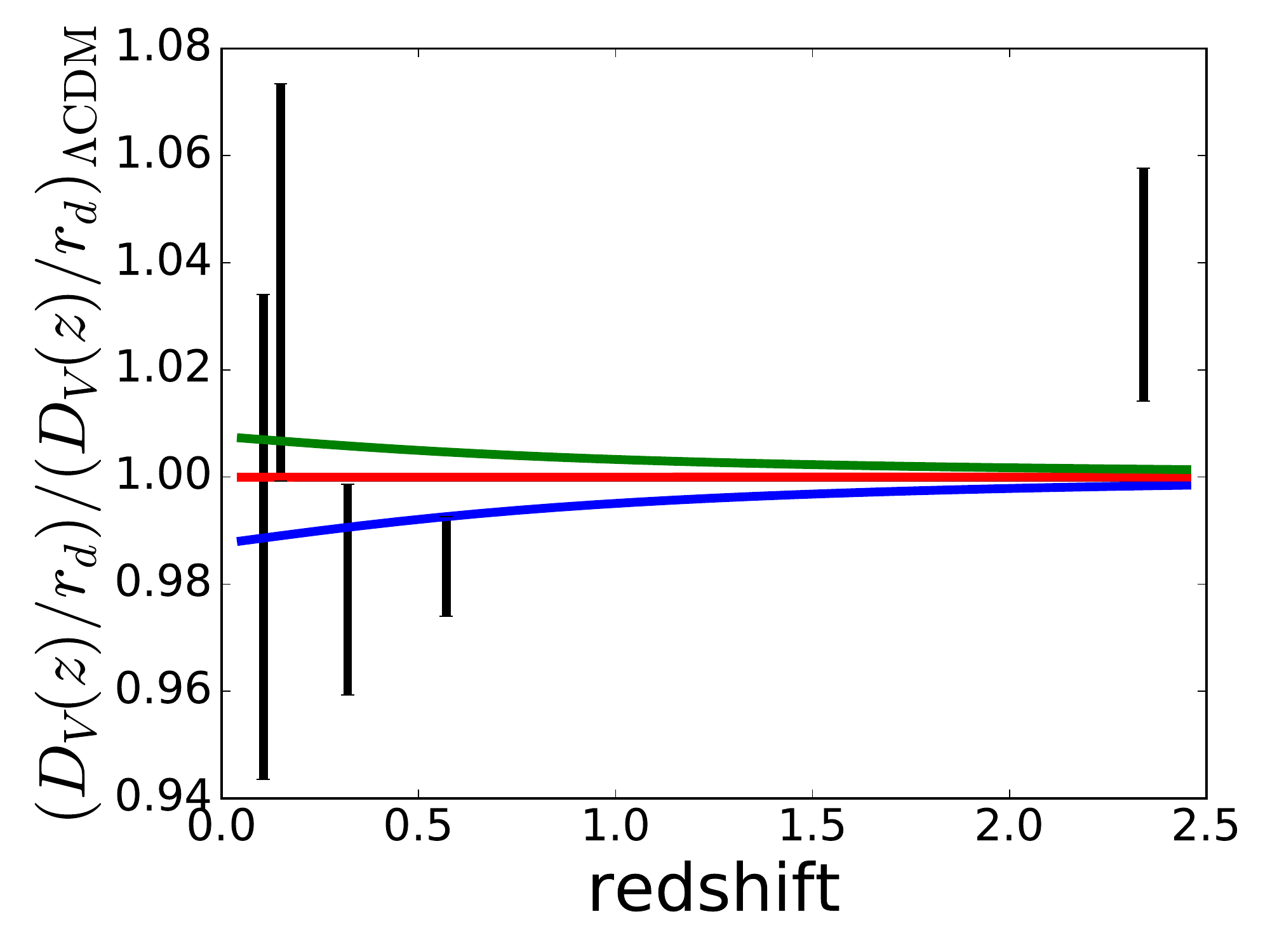}
\includegraphics[width=0.45\textwidth]{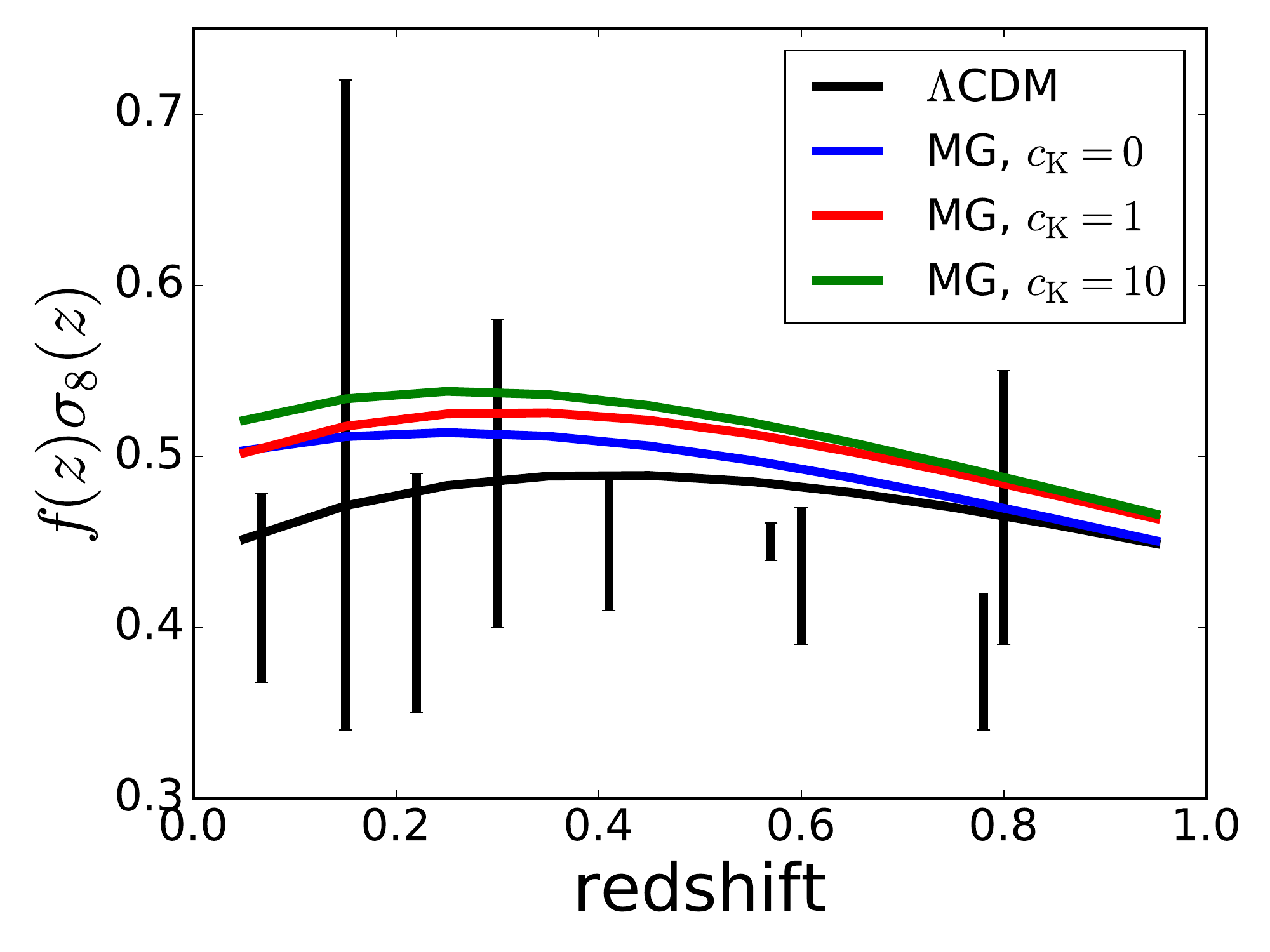}
\caption{Left panel: BAO measurements of the angle-averaged distance $D_V(z)/r_d$ used in this paper. Right panel: RSD measurements of the growth rate times the amplitude of fluctuations $f(z)\sigma_8(z)$ used in this paper. We also show the predictions from $\Lambda$CDM (black solid line) and MG best-fitting models to CMB only (blue lines: $c_\mathrm{K}=0$; red lines: $c_\mathrm{K}=1$; green lines: $c_\mathrm{K}=10$).\label{fig:data}}
\end{center}
\end{figure}

We use a compilation of recent cosmological data including CMB, large scale structure, redshift space distortions and baryon acoustic oscillations. To test the robustness of the results, different data sets combinations are analyzed, before proceeding to the global fit. 
\begin{itemize}
\item[CMB:]
We include the Cosmic Microwave Background (CMB) temperature power spectrum from \textit{Planck}. We compare three cases: we consider 1) Planck 2013 temperature power spectrum data \cite{Planck2013} with the low-$\ell$ multipoles of the polarization power spectrum from \textit{WMAP} \cite{Bennett:2012zja} (hereafter WP), 2) Planck 2015 temperature and polarization data \cite{Planck2015a,Planck2015b}, and 3) Planck 2015 temperature and polarization data with WMAP polarization replacing Planck polarization data at low multipoles (Planck2015WP). At the end we present our main results for our ``gold" data set Planck2015WP and refer the reader to the Appendix. We choose the latter as our default case because we find WP to be more constraining than Planck low-$\ell$ polarization for certain non-standard models (see Section~\ref{sec:results}). We emphasize that we use the full shape of the power spectrum rather than the shift parameters that determine the position of the acoustic peaks. This dataset includes about 7 $e$-folds out of the $\sim 60$ that we expect from inflation. This is the only probe we use from the high-$z$ Universe, much earlier than when late-time acceleration of the Universe took place. On the other hand, the multipoles below $\ell \lesssim 90$ probe super-horizon scales, which are relevant to this work, especially due to the anomalies reported in these scales (e.g., low value for the quadrupole \cite{Efstathiou2004,Slosar2004}, low value around $\ell \sim 20$ \cite{PlanckCl}, etc.).\\

\item[$P(k)$:]  We use the shape of the power spectrum of galaxies from the WiggleZ survey \cite{Blake2011} including scales in the range $10^{-4} h/\mathrm{Mpc} < k < 0.2 h/\mathrm{Mpc}$. These probe roughly another 7 $e$-folds. This constrains the shape of the power spectrum in the late Universe $0.2<z<1.0$ so this becomes sensitive to late-time effects. There are also claims of anomalies with these kind of datasets (e.g., \cite{BOSSneutrino}), mostly related with the value of the normalization of matter fluctuations $\sigma_8$ which is slightly different from the one derived from CMB data. Effects from massive neutrinos have also been invoked to bring these two datasets into better agreement. Uncertainties from modeling of the halo occupation distribution  that maps the relation between the galaxy and matter fluctuations have also been claimed \cite{Smith2015}.\\

\item [BAO:] Another cosmological probe we use is Baryon Acoustic Oscillations (BAO). BAO have their origin in the sound waves of the baryon-photon plasma in the early Universe, which were later imprinted in the distribution of galaxies, which we have observed with galaxy surveys such as 6dFGS \cite{Beutler2011}, BOSS \cite{Anderson2014, Tojeiro2014}, or WiggleZ \cite{Blake2011}. These show a feature at a characteristic scale of $\sim 150$ Mpc that is used to constrain the distance-redshift relation, effectively being used as a standard ruler. We note that these measurements tend to marginalize over the shape of the power spectrum (or correlation function) and focus on the position of the BAO peak, so they cannot constrain by themselves the matter power spectrum. We note that some anomalies have been reported regarding the BAO in the distribution of the Lyman-$\alpha$ forest, which seems to be in tension with the distance derived from the best-fit $\Lambda$CDM model derived from \textit{Planck} data. Since this BAO measurement corresponds to a redshift of $z=2.4$ in which we should not expect any influence from late-time acceleration effects, it is interesting to see if they affect our constraints on MG.
BAO measurements help fix the expansion history thus reducing possible parameter  degeneracies. The BAO measurements used are  summarized in Tab.~\ref{tab:bao} and shown in the left panel of Figure~\ref{fig:data}.\\

\item[RSD:]Finally we also include data from Redshift Space Distortions (RSD) in the galaxy distribution. These are also derived from Large Scale Structure surveys such as 6dFGS \cite{6DFRSD}, SDSS \cite{DR7RSD,MGSRSD}, BOSS \cite{CMASSRSDBETH}, WiggleZ \cite{WIGGLEZRSD}, or VIPERS \cite{VIPERSRSD}. Assuming linear theory, it is possible to use the anisotropies in the galaxy clustering at intermediate scales ($30h^{-1}-150h^{-1}$Mpc) to infer the line-of-sight contamination in the galaxy redshifts due to peculiar velocities in bulk motion \cite{Kaiser}. Similarly to (isotropic) galaxy clustering being able to constrain the combination $b\sigma_8$, where $b$ is the bias of the galaxy distribution with respect to the dark matter and $\sigma_8$ the amplitude of (linear) perturbations  on $8$Mpc/$h$ scales, RSD measures the combination $f\sigma_8$, where $f=d\log D/d\log a$ is known as the growth rate and is the logarithmic derivative of the growth factor. It is therefore a probe of the growth of structures and hence of MG. Since these measurements have been obtained with galaxy surveys below redshift $z\lesssim 1$, these should be affected by the same effects that caused the late-time acceleration of the Universe. Moreover, there have been claims that $f\sigma_8$ values from RSD are systematically low compared with those obtained from the best-fit $\Lambda$CDM model derived from \textit{Planck} data \cite{Planck2015b}, which makes this dataset very relevant for this study. The redshift space distortions data compilation  we adopt is summarized in Tab.~\ref{tab:rsd} and shown in the right panel of Figure~\ref{fig:data}.\\
\end{itemize}

\begin{table}
\begin{center}
\begin{tabular}{cccc}
\hline
Survey     & Measurement        & Value           & Reference \\ 
\hline
6dFGS      & $r_d/D_V(z=0.106)$ & 0.327$\pm$0.015 & \cite{Beutler2011} \\
SDSS-MGS   & $D_V(z=0.15)/r_d$  & 4.47$\pm$0.16   & \cite{Ross2014}    \\
BOSS-LOWZ  & $D_V(z=0.32)/r_d$  & 8.47$\pm$0.17   & \cite{Tojeiro2014} \\
BOSS-CMASS & $D_V(z=0.57)/r_d$  & 13.77$\pm$0.13  & \cite{Anderson2014}\\
BOSS-LYA   & $D_A(z=2.36)/r_d$  & 10.8$\pm$0.4    & \cite{Font2014} \\
BOSS-LYA   & $c/H(z=2.36)r_d$   & 9.0$\pm$0.3     & \cite{Font2014} \\
\hline
\end{tabular}
\caption{List of the BAO measurements used in this paper. \label{tab:bao}}
\end{center}
\end{table}

\begin{table}
\begin{center}
\begin{tabular}{cccc}
\hline
Survey     & $z$    & $f(z)\sigma_8(z)$ & Reference \\
\hline
6dFGS      & 0.067  & 0.423$\pm$0.055   & \cite{6DFRSD} \\
MGS        & 0.15   & 0.53$\pm$0.19     & \cite{MGSRSD} \\
LRG        & 0.30   & 0.49$\pm$0.09     & \cite{DR7RSD} \\
WIGGLEZ    & 0.22   & 0.42$\pm$0.07     & \cite{WIGGLEZRSD} \\
WIGGLEZ    & 0.41   & 0.45$\pm$0.04     & \cite{WIGGLEZRSD} \\
WIGGLEZ    & 0.60   & 0.43$\pm$0.04     & \cite{WIGGLEZRSD} \\
WIGGLEZ    & 0.78   & 0.38$\pm$0.04     & \cite{WIGGLEZRSD} \\
BOSS       & 0.57   & 0.450$\pm$0.011   & \cite{CMASSRSDBETH} \\
VIPERS     & 0.80   & 0.47$\pm$0.08     & \cite{VIPERSRSD} \\
\hline
\end{tabular}
\caption{List of the RSD measurements used in this paper. \label{tab:rsd}}
\end{center}
\end{table}

\section{Results}
\label{sec:results}

We now present our constraints on the Horndeski parameters derived from Monte Carlo sampling of this parameter space using the cosmological datasets described in the previous Section. For CMB data we use the compilation Planck2015WP. 
We have also produced constraints using Planck 2013+WP and  using Planck 2015 data \citep{Planck2015a, Planck2015b}, which uses Planck's own polarization measurements instead of 
the low-$\ell$ polarization data from WMAP.  Because of the longer integration time and the additional information from the high $\ell$ polarization offered by Planck 2015, the  na{\"i}ve expectation is that Planck 2015 should be more constraining than   Planck 2013.
If we enforce a hard prior on the integrated optical depth due to reionization $\tau_{\rm reio} >0.04$ that is indeed the case. However,  our results  depend on the choice 
of the prior. We conclude that for the non-standard models considered here the low $\ell$ polarization measurements from WMAP are more constraining that Planck's.  This is why  we choose to report our updated constraints using as the CMB dataset a combination of the high multipoles of Planck temperature and polarization power spectrum, the low multipoles of Planck temperature power spectrum, and the low multipoles of WMAP polarization power spectrum. A  detailed comparison between constraints from these different  CMB datasets is presented in the Appendix.

First of all, we note that the coefficient $c_\textrm{K}$ is largely unconstrained by the data, even when the sampling is done in logarithmic space. This was expected, since $\alpha_\textrm{K}$ does not introduce new scale dependence and it does not modify the linear growth. In the quasi-static approximation  where time derivatives are considered to be sub-dominant with respect to space derivatives, $\alpha_\mathrm{K}$ does not enter the equations of motion. In this regime we expect no effect of $\alpha_\mathrm{K}$ on the observables and thus no sensitivity of our analysis to this parameter.
Therefore, we choose a few arbitrary  fixed values of  $c_\textrm{K}$  and report constraints for these cases. 
In Table \ref{tab:alphas} we report the 95\% confidence level bounds on the values of $c_\textrm{B,M,T}$ using different combinations of datasets and after marginalizing over all the other parameters in the MCMC chain. In interpreting this table one should bear in mind that  the case $c_\textrm{K}=c_\textrm{B}=c_\textrm{M}=c_\textrm{T}=0$ corresponds to the $\Lambda$CDM model.

The table shows that there is some mild dependence of the constraints on the adopted value of $c_\textrm{K}$.
In all cases (all data set combinations and all values of $c_\textrm{K}$)  the parameter  $c_\textrm{M}$ is consistent with zero. For this parameter the  $c_\textrm{K}$ dependence is very mild.

When RSD is not included, the $c_\textrm{T}$ parameter is consistent with zero and shows no $c_\textrm{K}$ dependence for the data set combinations CMB, CMB+BAO and CMB+PK. For the CMB+RSD and CMB+BAO+PK+RSD data sets this parameter is negative and 0 is formally outside the 95\% confidence. The $c_\textrm{T}$ lower limit is imposed by the prior $c_\textrm{T}>-0.9$. It is then clear that RSD pushes $c_\textrm{T}$ towards negative values. Indeed, of the data we consider RSD is the only one that probes (with good statistical  significance) the growth of  perturbations (in fact it measures $f\sigma_8$). There is some small dependence of that in CMB lensing and ISW but at lower significance. It is well known that within the LCDM paradigm,  RSD wants an amplitude of fluctuations slightly lower than that predicted  when fitting the model to CMB data.  Negative values of the parameter  $c_\textrm{T}$ achieve just that: a lower $f\sigma_8$ at the redshift of interest for RSD without major alterations to other observables.

The  confidence region for the parameter  $c_\textrm{B}$  also excludes $0$ in few cases and positive values seems to be formally  favoured. The upper limit of  $c_\textrm{B}$ does not depend on the adopted value of $c_\textrm{K}$, but it is sensitive to the choice of data set. The lower limit on the other hand is also sensitive to the adopted $c_\textrm{K}$ value.
 
Our tightest constraints, which include all the datasets considered, 
formally appear to favour a theory of gravity that is minimally coupled with tensors propagating slower than the speed of light. The results are robust with respect to changes in the low-$\ell$ CMB polarization data (from \textit{Planck} to \textit{WMAP}, see Appendix~\ref{sec:appendix}). We have also tested that the model-dependent RSD measurement in \cite{CMASSRSDBETH} is not driving our results, by checking the consistency of the constraints when that data point is replaced with the measurement in \cite{CMASSRSD}. These tests confirm the robustness of our results.

\begin{table}
\begin{center}
\begin{tabular}{lccc}
                         & $c_\textrm{B}$         & $c_\textrm{M}$         & $c_\textrm{T}$         \\
\hline
CMB, $c_\textrm{K}=0$         & $+0.04<c_\textrm{B}<+1.91$ & $-0.86<c_\textrm{M}<+2.00$ & $-0.90<c_\textrm{T}<+1.20$ \\
CMB, $c_\textrm{K}=1$         & $-0.01<c_\textrm{B}<+1.84$ & $-0.77<c_\textrm{M}<+2.00$ & $-0.90<c_\textrm{T}<+1.08$ \\
CMB, $c_\textrm{K}=10$        & $+0.14<c_\textrm{B}<+1.92$ & $-0.73<c_\textrm{M}<+2.00$ & $-0.90<c_\textrm{T}<+0.77$ \\
\hline
CMB+BAO, $c_\textrm{K}=0$     & $+0.07<c_\textrm{B}<+1.95$ & $-0.83<c_\textrm{M}<+2.00$ & $-0.90<c_\textrm{T}<+1.14$ \\
CMB+BAO, $c_\textrm{K}=1$     & $+0.08<c_\textrm{B}<+1.96$ & $-0.85<c_\textrm{M}<+2.00$ & $-0.90<c_\textrm{T}<+1.21$ \\
CMB+BAO, $c_\textrm{K}=10$    & $+0.19<c_\textrm{B}<+1.97$ & $-0.79<c_\textrm{M}<+2.00$ & $-0.90<c_\textrm{T}<+0.87$ \\
\hline
CMB+RSD, $c_\textrm{K}=0$     & $+0.27<c_\textrm{B}<+2.48$ & $-1.43<c_\textrm{M}<-0.23$ & $-0.90<c_\textrm{T}<-0.44$ \\
CMB+RSD, $c_\textrm{K}=1$     & $+0.15<c_\textrm{B}<+2.37$ & $-1.39<c_\textrm{M}<-0.13$ & $-0.90<c_\textrm{T}<-0.44$ \\
CMB+RSD, $c_\textrm{K}=10$    & $+0.20<c_\textrm{B}<+2.39$ & $-1.40<c_\textrm{M}<-0.14$ & $-0.90<c_\textrm{T}<-0.44$ \\
\hline
CMB+PK, $c_\textrm{K}=0$     & $+0.00<c_\textrm{B}<+1.82$ & $-0.85<c_\textrm{M}<+2.00$ & $-0.90<c_\textrm{T}<+0.99$ \\
CMB+PK, $c_\textrm{K}=1$     & $-0.03<c_\textrm{B}<+1.85$ & $-0.83<c_\textrm{M}<+2.00$ & $-0.90<c_\textrm{T}<+0.88$ \\
CMB+PK, $c_\textrm{K}=10$    & $+0.12<c_\textrm{B}<+1.87$ & $-0.80<c_\textrm{M}<+2.00$ & $-0.90<c_\textrm{T}<+0.72$ \\
\hline
CMB+BAO+RSD+PK, $c_\textrm{K}=0$ & $+0.24<c_\textrm{B}<+2.32$ & $-1.36<c_\textrm{M}<-0.13$ & $-0.90<c_\textrm{T}<-0.39$ \\
CMB+BAO+RSD+PK, $c_\textrm{K}=1$ & $+0.10<c_\textrm{B}<+2.29$ & $-1.35<c_\textrm{M}<-0.08$ & $-0.90<c_\textrm{T}<-0.41$ \\
CMB+BAO+RSD+PK, $c_\textrm{K}=10$ & $+0.19<c_\textrm{B}<+2.30$ & $-1.36<c_\textrm{M}<-0.06$ & $-0.90<c_\textrm{T}<-0.41$ \\
\hline
\end{tabular}
\caption{Constraints on the coefficients $c_\textrm{B}$, $c_\textrm{M}$, and $c_\textrm{T}$ from different cosmological dataset combinations and for different values of $c_\textrm{K}$. Quoted limits are 95\% CL. A hard prior on $c_\textrm{T}>-0.9$ is applied  as well as a prior on $-2<c_M<+2$ that has become relevant in some cases.\label{tab:alphas}}
\end{center}
\end{table}

For the other data sets the  $\Lambda$CDM model ($c_\textrm{K}=c_\textrm{B}=c_\textrm{M}=c_\textrm{T}=0$) is not always included in the 95\% posterior confidence interval.

We will study below the Bayesian evidence ratio  of the $\Lambda$CDM to  MG models. But before we do so, 
in order to interpret the  significance of this result we examine carefully the values of the {\it effective} $\chi^2$ statistic --given by the log of the likelihood-- measured at the best-fit point in our Monte Carlo chains.  This is shown in Table \ref{tab:loglike}, where we also include the standard $\Lambda$CDM case (in which the MG equations are not used) for direct comparison. We can see that CMB data actually prefers a larger value of the coefficient $c_\textrm{K}$, and with a lower significance this is also true for RSD data. On the other hand, PK data slightly compensate this trend by selecting lower values of $c_\textrm{K}$ at the best-fit model.

The MCMC procedure is not optimized to find the best fit model which maximizes the likelihood, therefore there is an intrinsic error associated to these numbers which have been estimated to be $\sim 0.7 $\cite{Verde2013}.
 
Compared to the $\Lambda$CDM model, we find that the improvement on the fitting of cosmological data due to  the extra degrees of freedom provided by the Horndeski parameters is  not significant in most of the cases. Possible exception are the CMB (all  $c_\textrm{K}$ values)  and the CMB+BAO for $c_\textrm{K}=10$ cases where the improvement is $\Delta$ log Likelihood $\sim 2$ at the ``cost" of three extra degrees of freedom.
This suggests that the deviations found in our datasets are still consistent with a fluctuation within the $\Lambda$CDM scenario, even though (remarkably) the posterior distributions of  MG coefficients presented in Table~\ref{tab:alphas} are not  always consistent with zero.

\begin{table}
\small
\begin{tabular}{lccccccc}
 - log likelihood                      & CMB  & CMB  & BAO & RSD & PK & Total &$\Delta^{\rm{Total}}_{\Lambda CDM}$ \\
                    &  (low-$\ell$) &  (high-$\ell$) &  &  &  &  & \\
\hline
CMB ($\Lambda$CDM)                     & 1014.7 & 1222.2 &     &     &       & 2236.9 & -- \\
CMB (MG), $c_\textrm{K}=0$             & 1014.4 & 1221.9 &     &     &       & 2236.3 & 0.6\\
CMB (MG), $c_\textrm{K}=1$             & 1014.3 & 1221.8 &     &     &       & 2236.1 & 0.8\\
CMB (MG), $c_\textrm{K}=10$            & 1013.5 & 1221.4 &     &     &       & 2234.9 & 2.0\\
\hline
CMB+BAO ($\Lambda$CDM)                 & 1015.1 & 1221.6 & 5.6 &     &       & 2242.3 & -- \\
CMB+BAO (MG), $c_\textrm{K}=0$         & 1014.8 & 1221.8 & 5.0 &     &       & 2241.5 & 0.8\\
CMB+BAO (MG), $c_\textrm{K}=1$         & 1013.9 & 1221.1 & 5.2 &     &       & 2240.2 & 2.1\\
CMB+BAO (MG), $c_\textrm{K}=10$        & 1014.0 & 1221.7 & 4.8 &     &       & 2240.5 & 1.8\\
\hline
CMB+RSD ($\Lambda$CDM)                 & 1014.3 & 1223.7 &     & 4.6 &       & 2242.6 & -- \\
CMB+RSD (MG), $c_\textrm{K}=0$         & 1013.7 & 1222.1 &     & 2.2 &       & 2237.9 & 4.7\\
CMB+RSD (MG), $c_\textrm{K}=1$         & 1013.9 & 1223.3 &     & 2.1 &       & 2239.3 & 3.3\\
CMB+RSD (MG), $c_\textrm{K}=10$        & 1013.4 & 1222.3 &     & 2.7 &       & 2238.4 & 4.2\\
\hline
CMB+PK ($\Lambda$CDM)                  & 1015.3 & 1221.5 &     &     & 228.6 & 2465.5 & -- \\
CMB+PK (MG), $c_\textrm{K}=0$          & 1014.1 & 1222.6 &     &     & 229.8 & 2466.5 & {\color{red}-1.0}\\
CMB+PK (MG), $c_\textrm{K}=1$          & 1013.8 & 1222.4 &     &     & 229.2 & 2465.4 & 0.1\\
CMB+PK (MG), $c_\textrm{K}=10$         & 1013.3 & 1221.9 &     &     & 229.5 & 2464.7 & 0.8\\
\hline
CMB+BAO+RSD+PK ($\Lambda$CDM)          & 1014.5 & 1226.5 & 4.7 & 3.1 & 228.2 & 2477.0 & --\\
CMB+BAO+RSD+PK (MG), $c_\textrm{K}=0$  & 1013.5 & 1222.9 & 4.8 & 1.9 & 229.0 & 2472.1 & 4.9\\
CMB+BAO+RSD+PK (MG), $c_\textrm{K}=1$  & 1013.7 & 1222.9 & 4.9 & 1.8 & 228.9 & 2472.2 & 4.8\\
CMB+BAO+RSD+PK (MG), $c_\textrm{K}=10$ & 1013.7 & 1222.7 & 4.8 & 1.7 & 228.9 & 2471.8 & 5.2\\
\hline
\end{tabular}
\caption{Absolute value of the log likelihoods (i.e. $\chi^2/2$) at the best fit point from the individual data that comprises each dataset combination explored in our analysis. The column labelled \textit{Total} displays the maximum likelihood value in the chain. The last column shows the difference in  Log likelihood with respect to the $\Lambda$CDM model. Red (negative) numbers represent worst fit, positive (black) numbers  better fit. Given the intrinsic uncertainty of the MCMC in determining the  best likelihood value, the improvement in $\chi^2$  offered by the more complex model is in most cases not significant. \label{tab:loglike}}
\end{table}

\begin{figure}
\begin{center}
\includegraphics[width=0.3\textwidth]{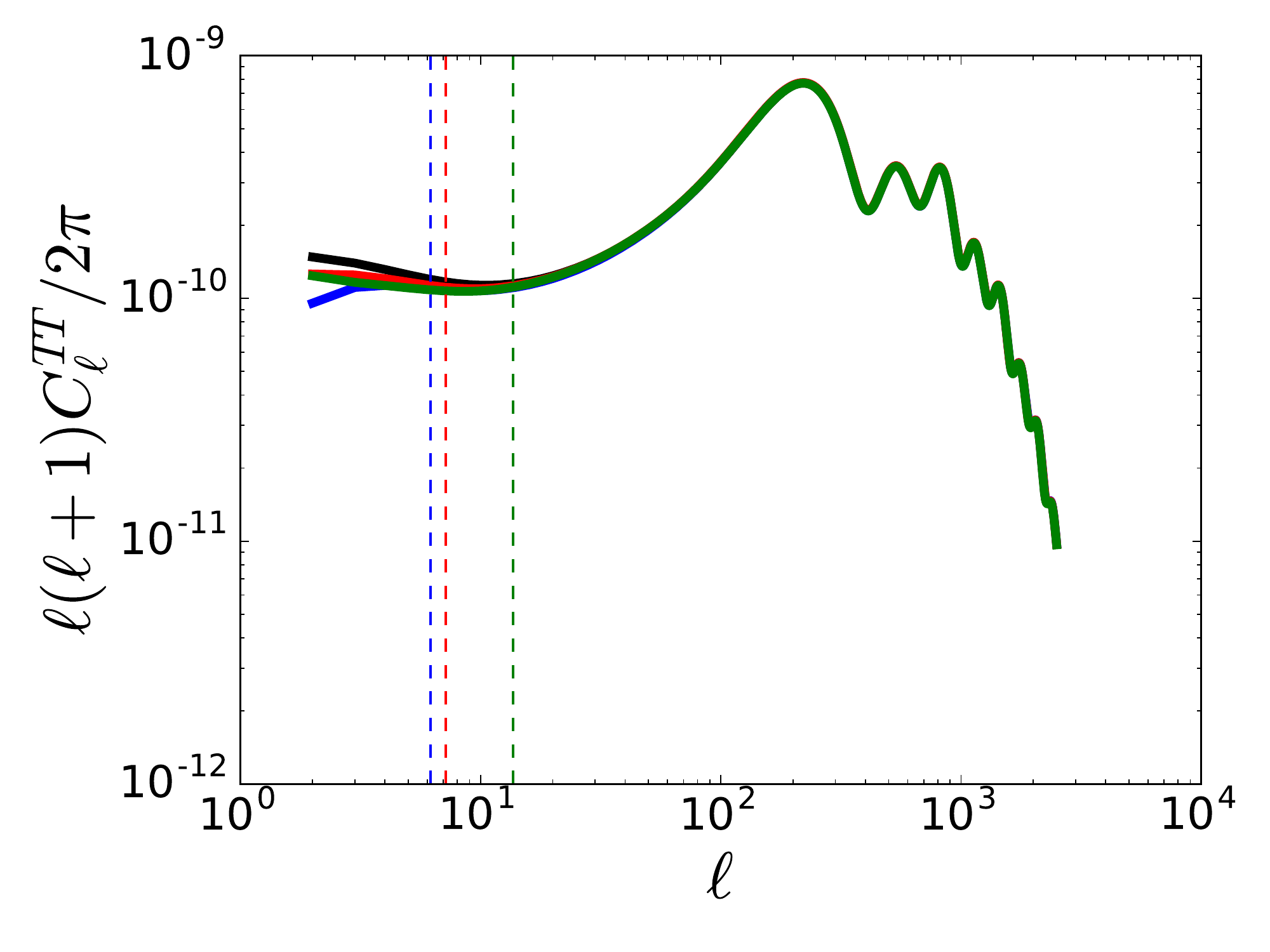}
\includegraphics[width=0.3\textwidth]{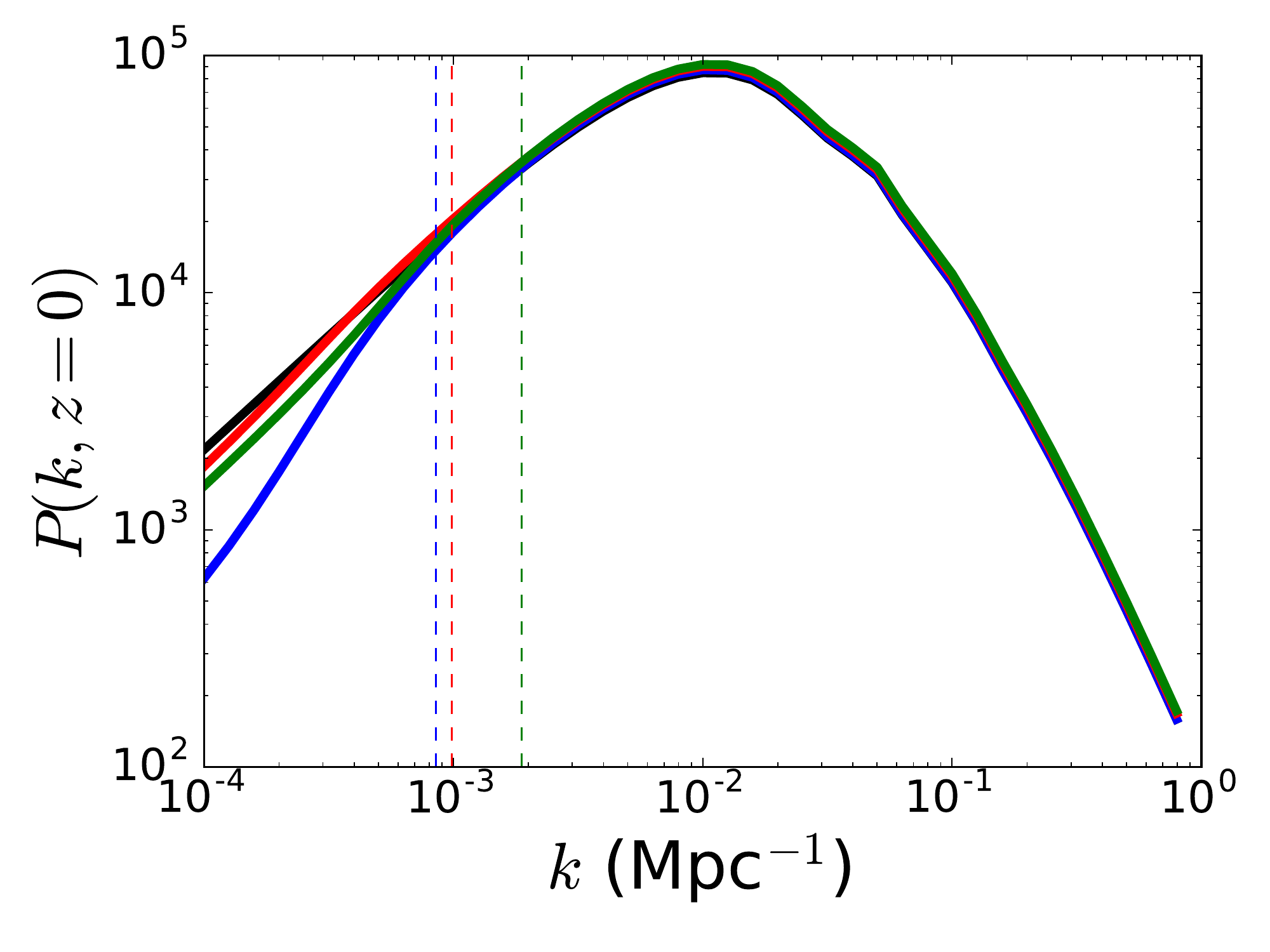}
\includegraphics[width=0.3\textwidth]{plots/cmbonly_lowl_rsd.pdf}
\includegraphics[width=0.3\textwidth]{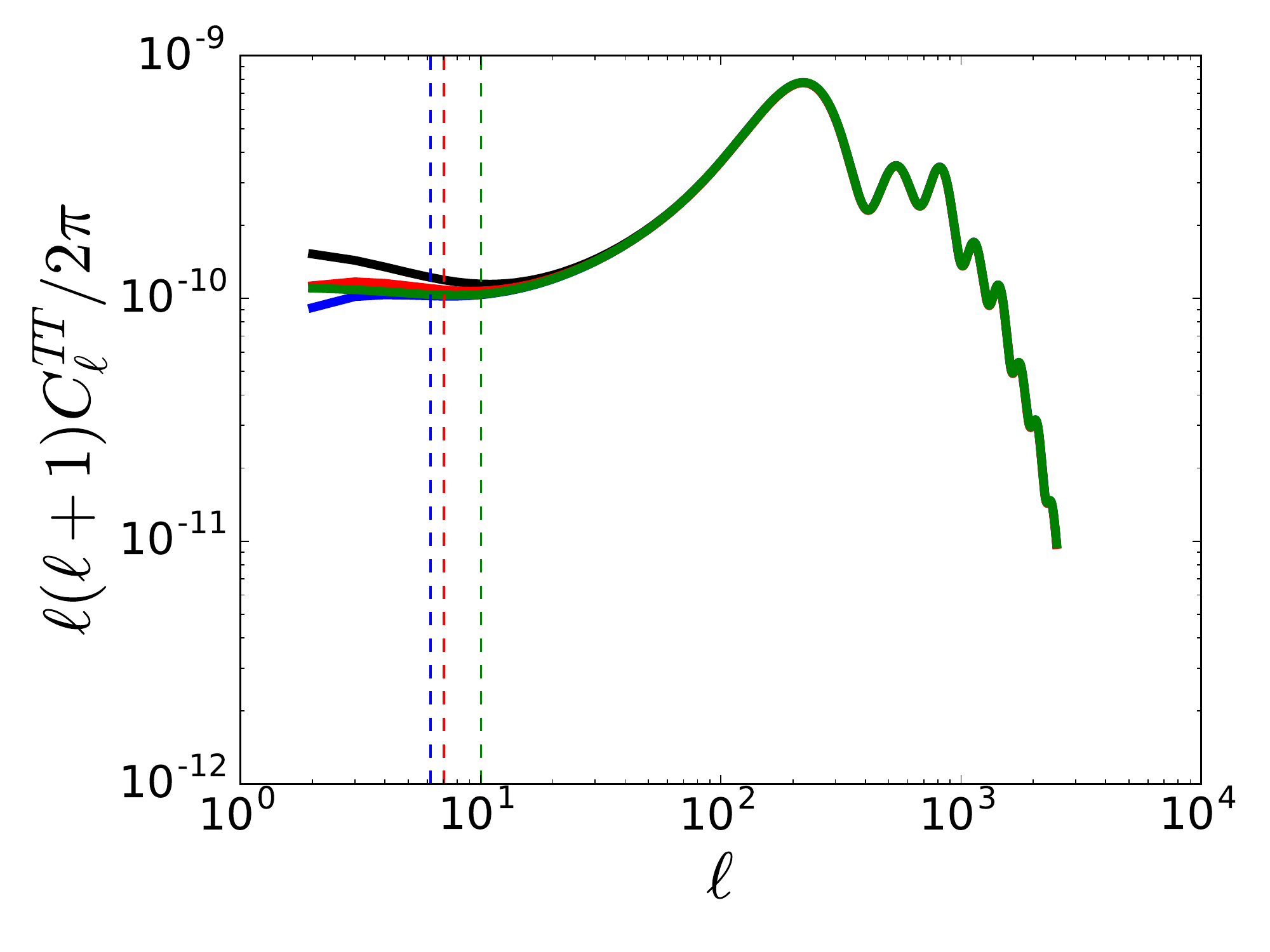}
\includegraphics[width=0.3\textwidth]{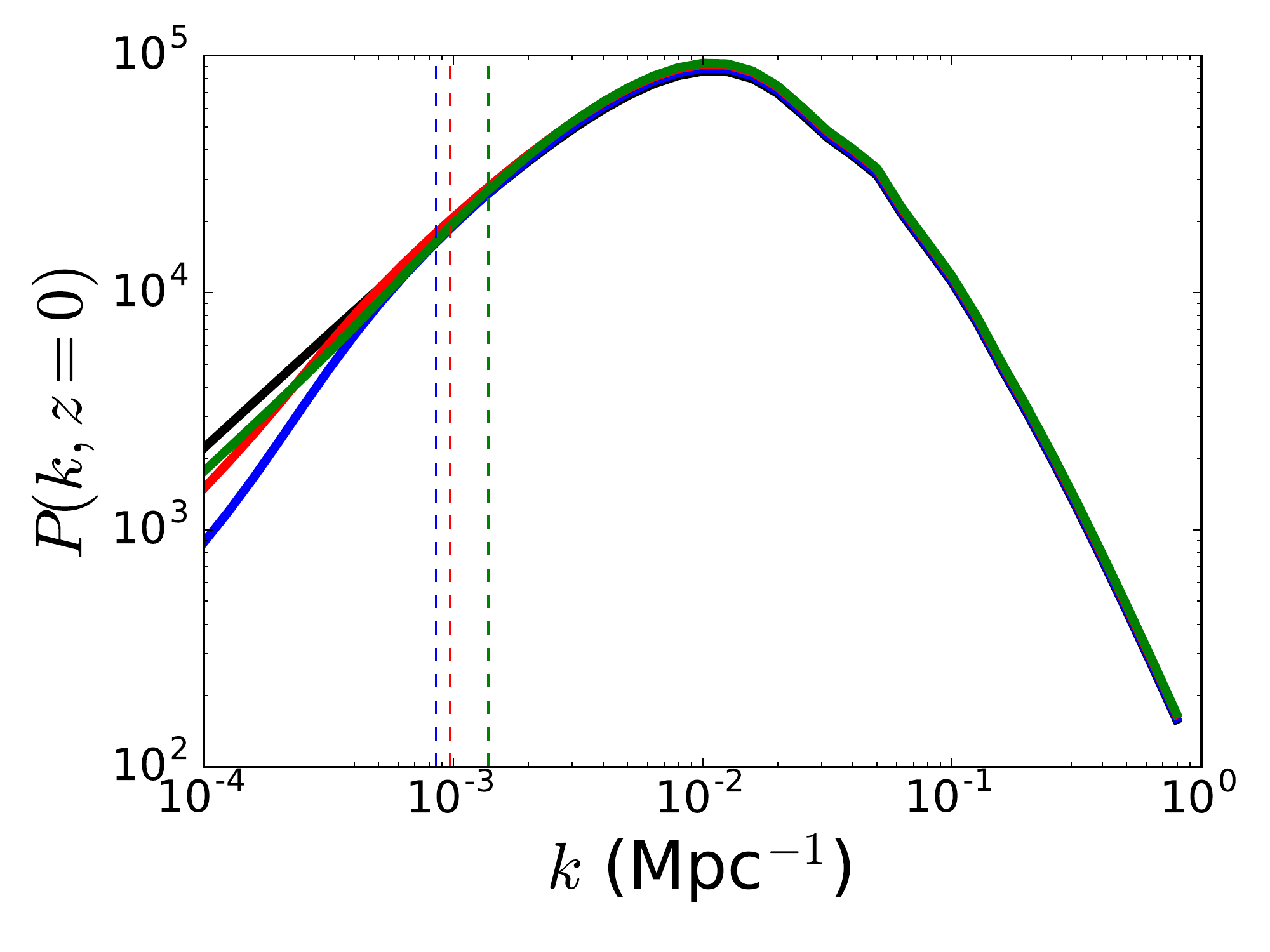}
\includegraphics[width=0.3\textwidth]{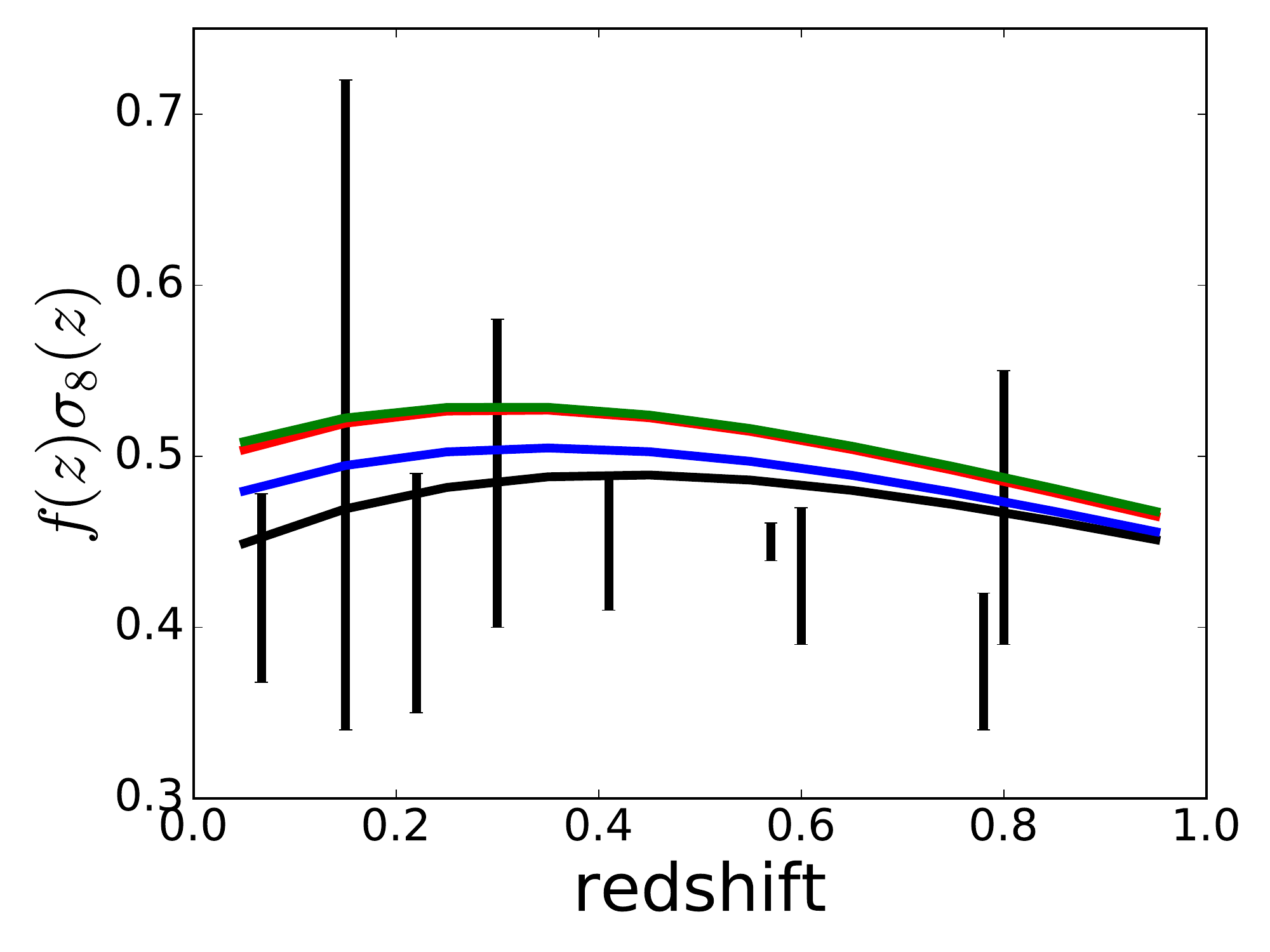}
\includegraphics[width=0.3\textwidth]{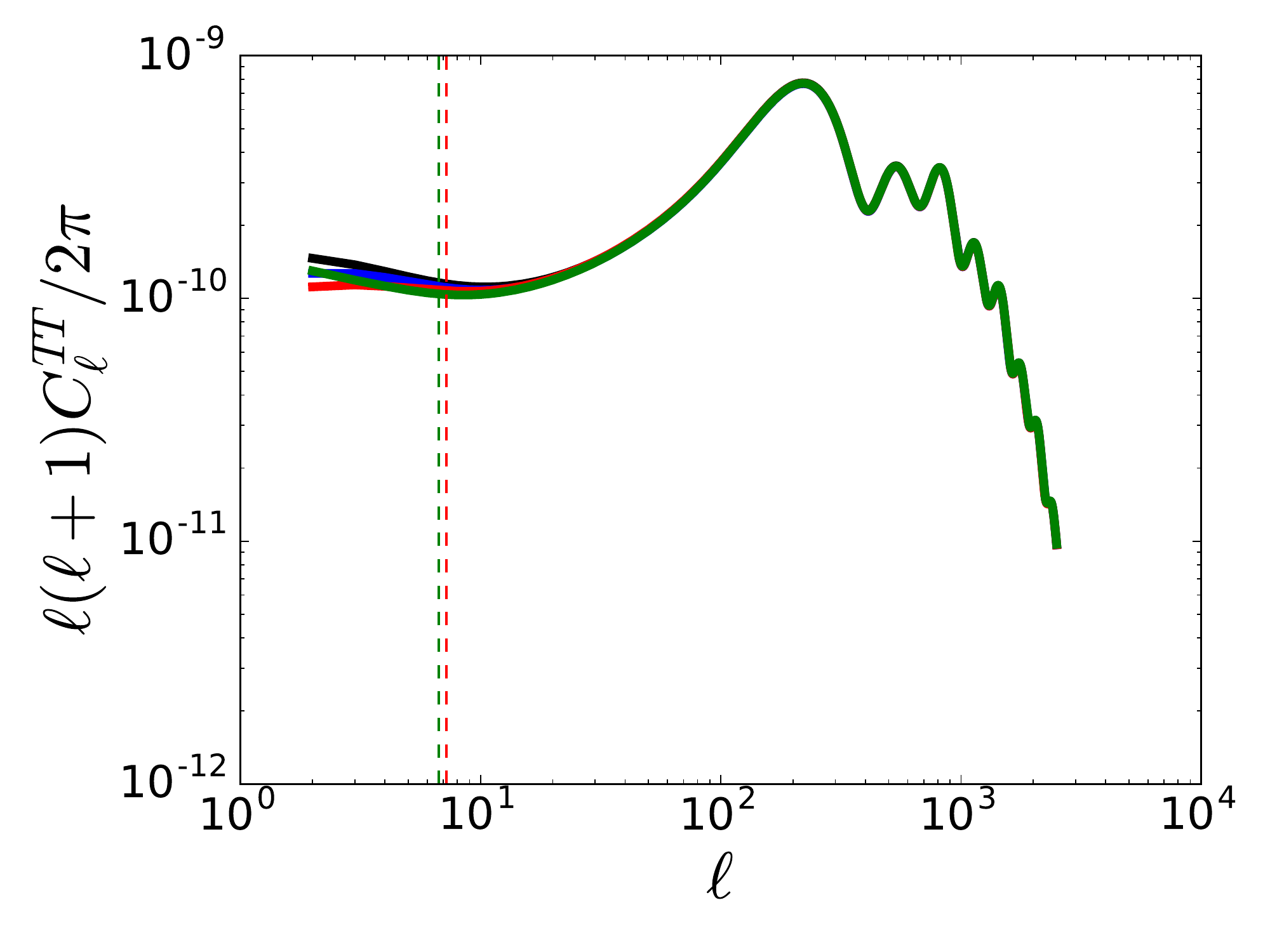}
\includegraphics[width=0.3\textwidth]{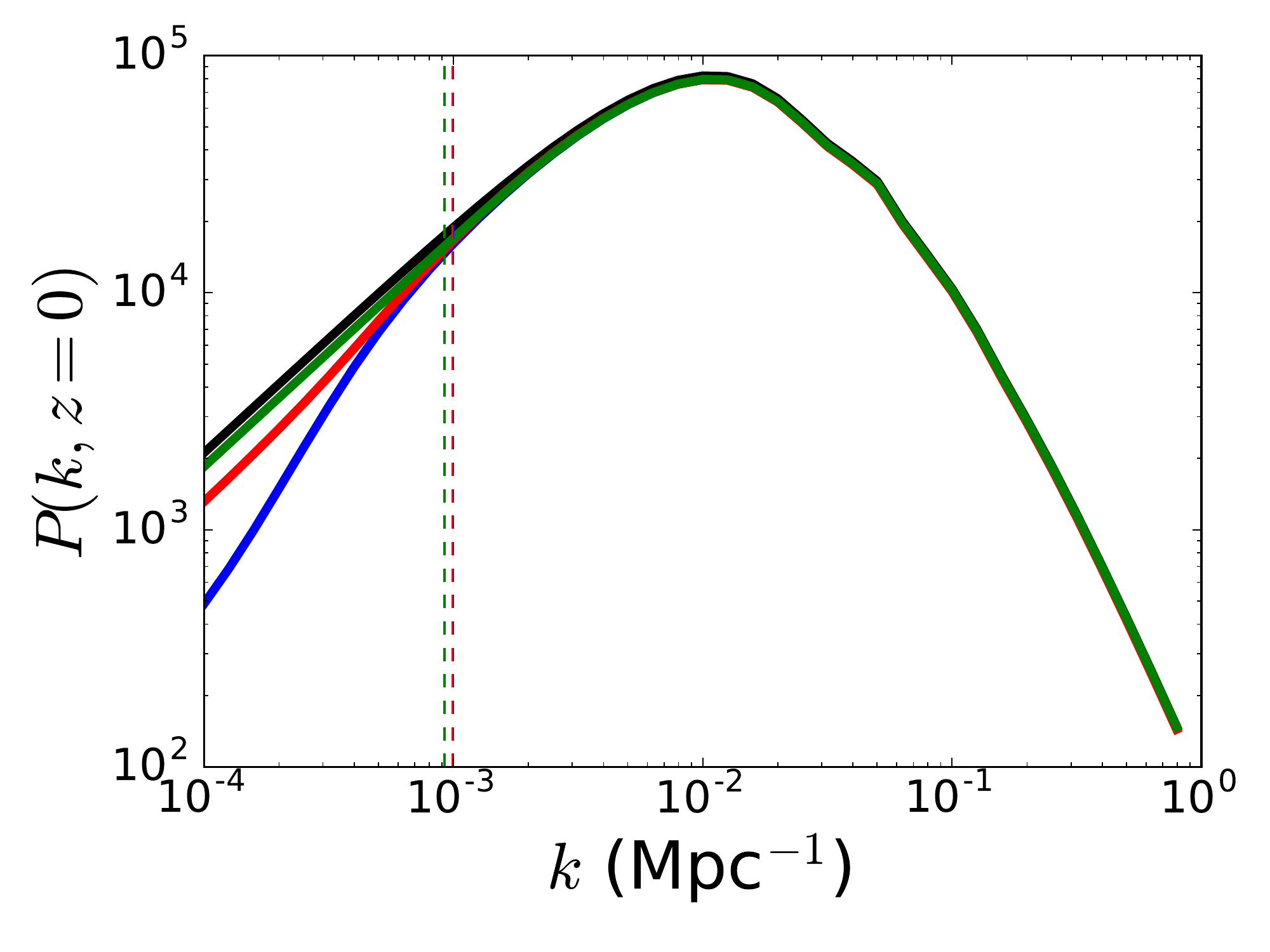}
\includegraphics[width=0.3\textwidth]{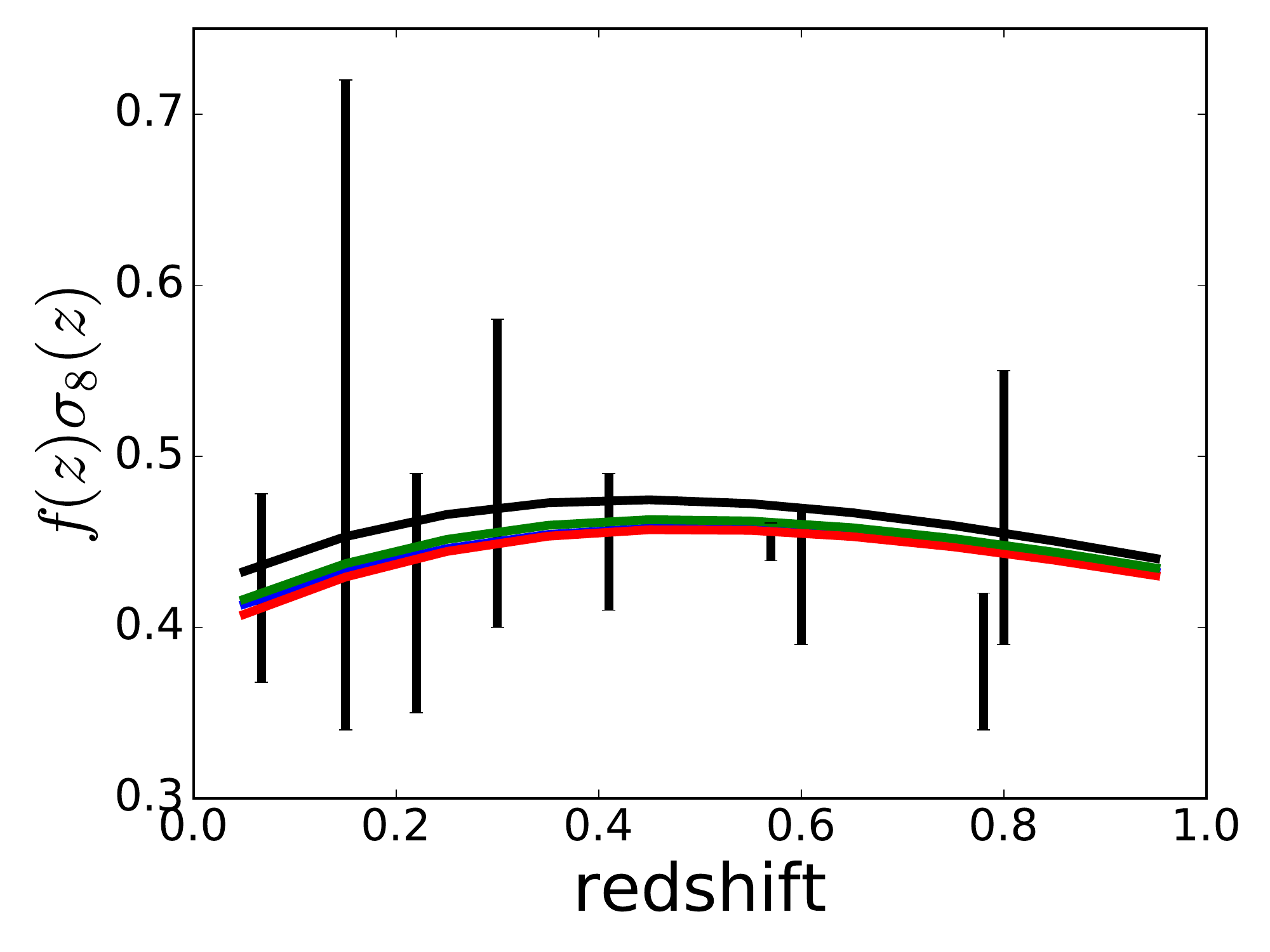}
\includegraphics[width=0.3\textwidth]{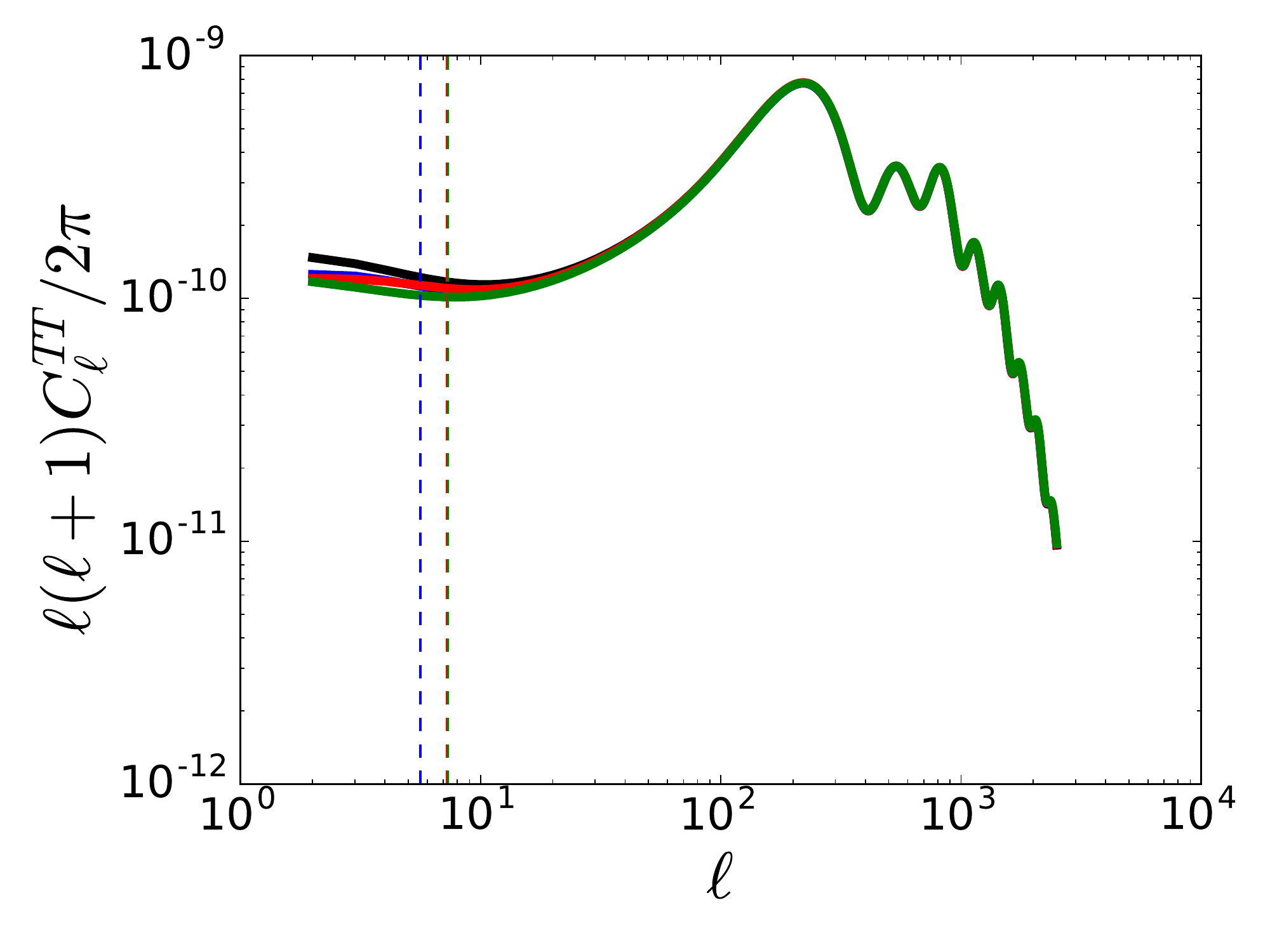}
\includegraphics[width=0.3\textwidth]{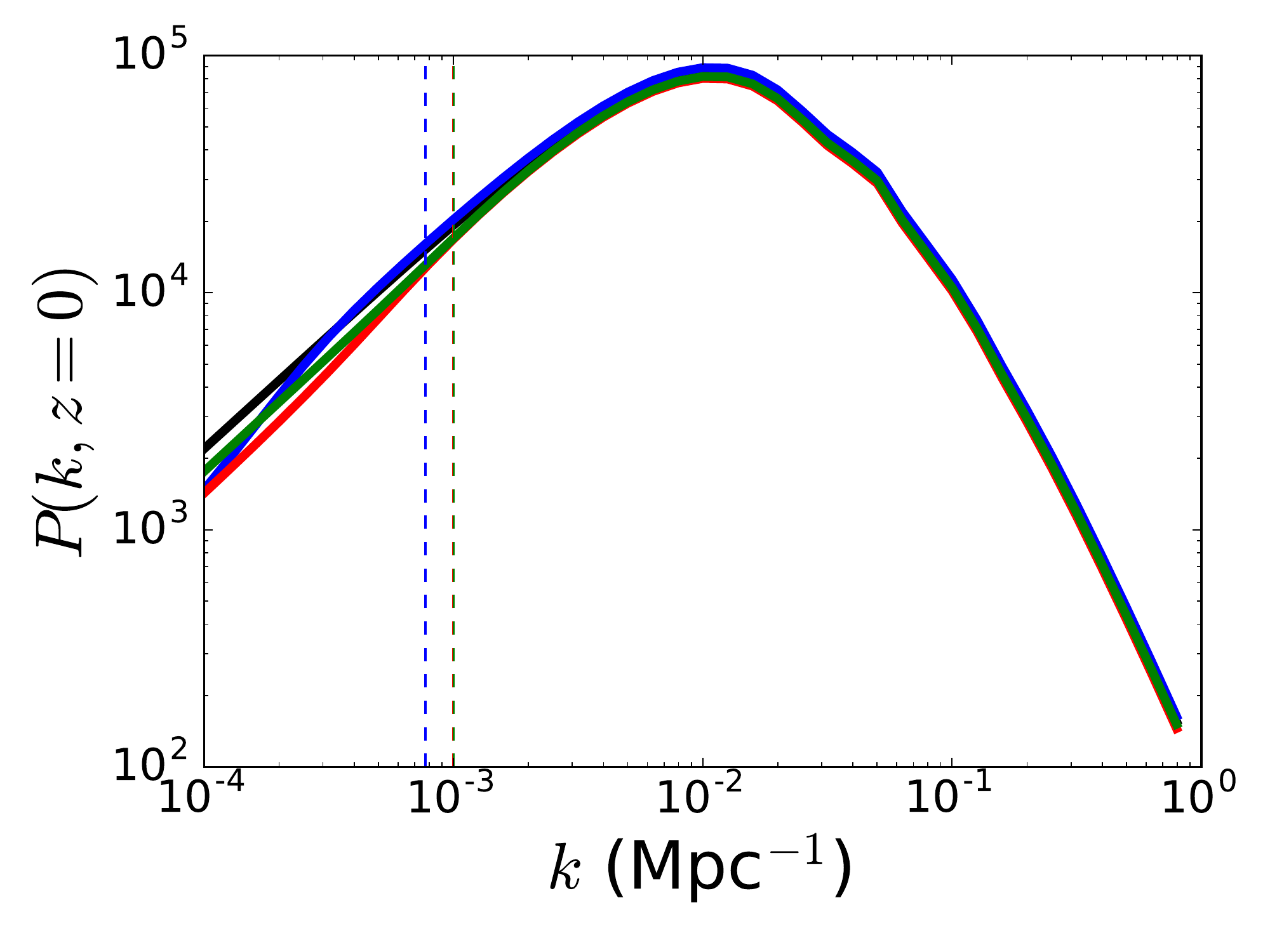}
\includegraphics[width=0.3\textwidth]{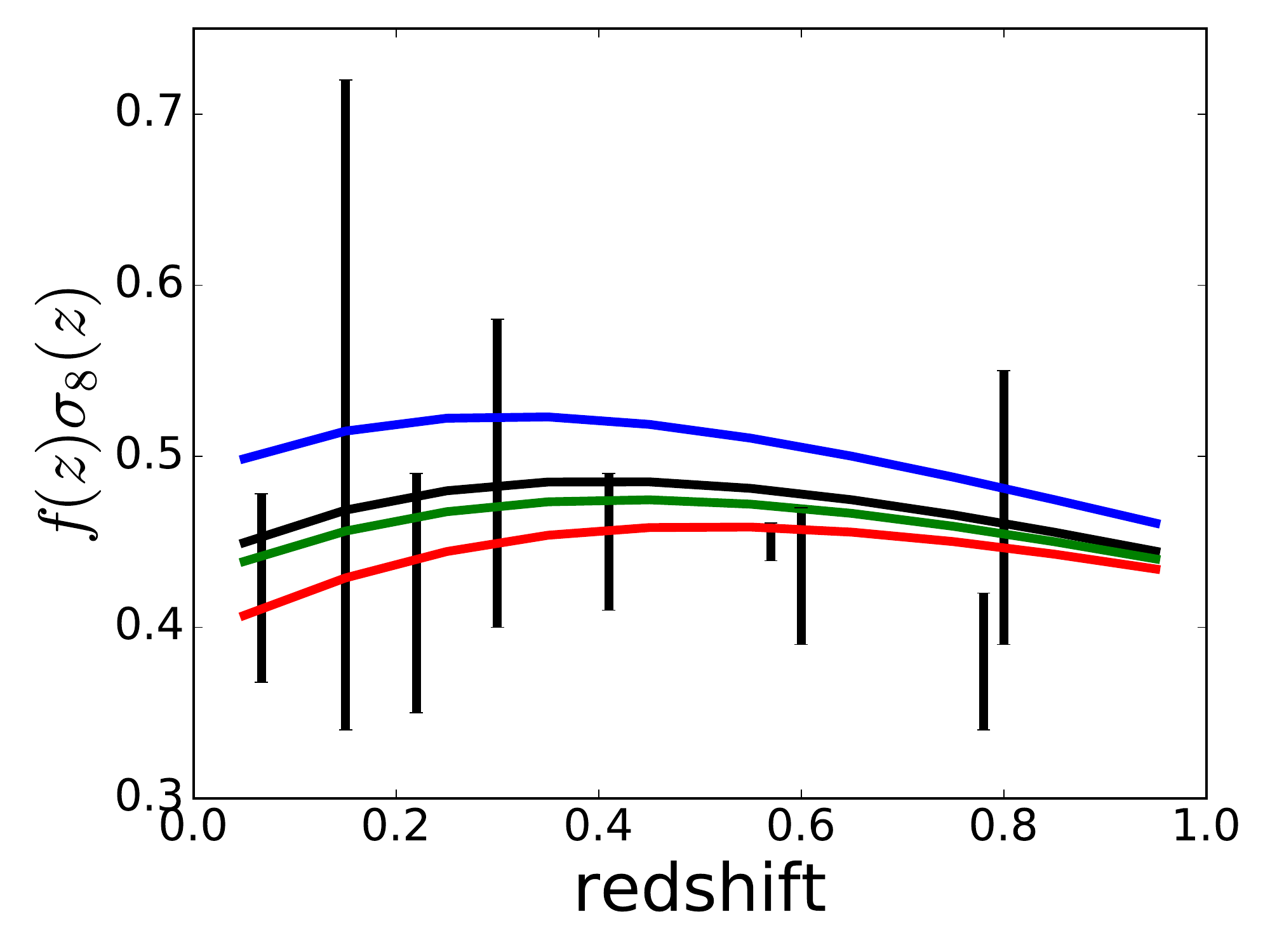}
\includegraphics[width=0.3\textwidth]{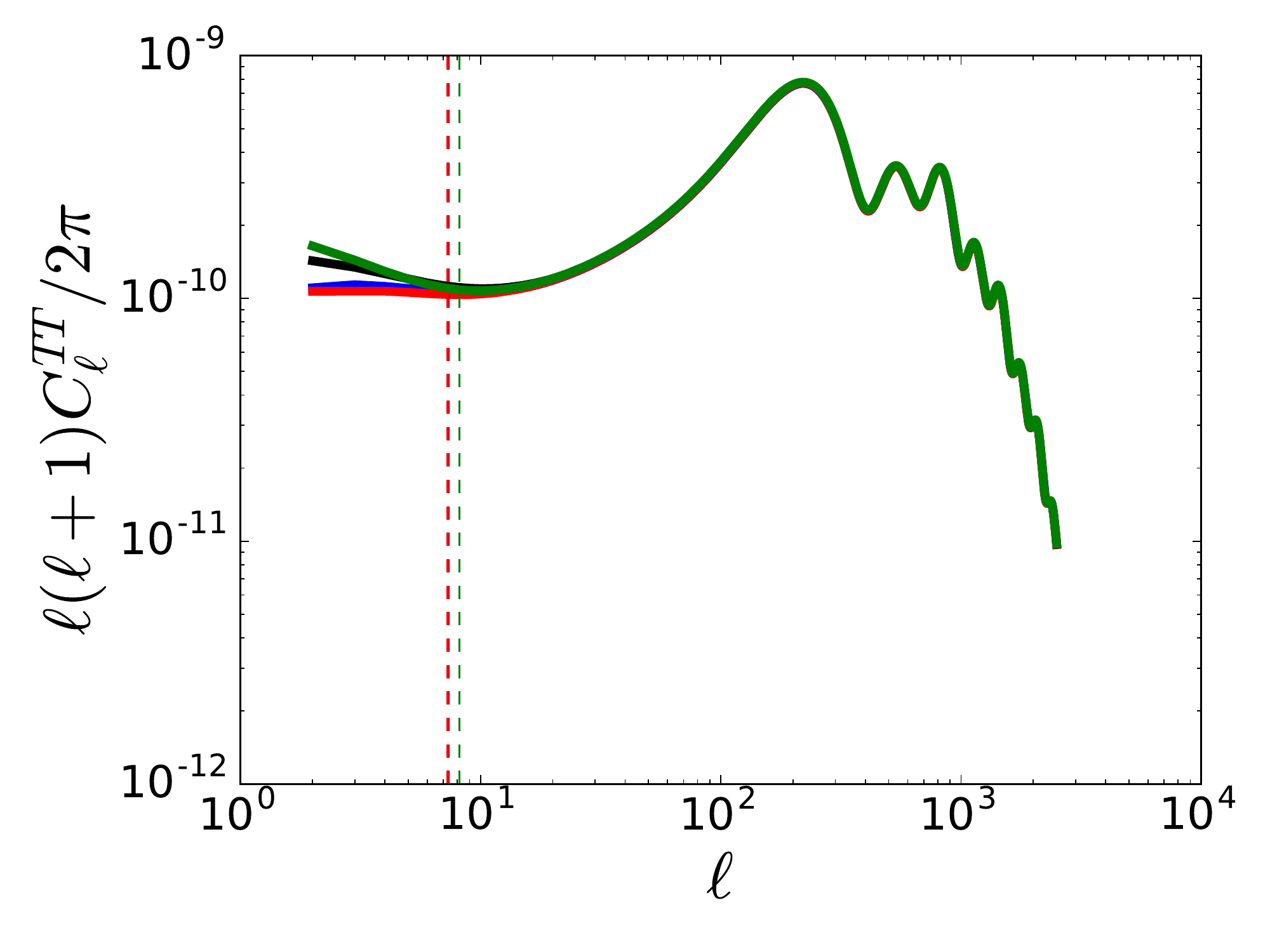}
\includegraphics[width=0.3\textwidth]{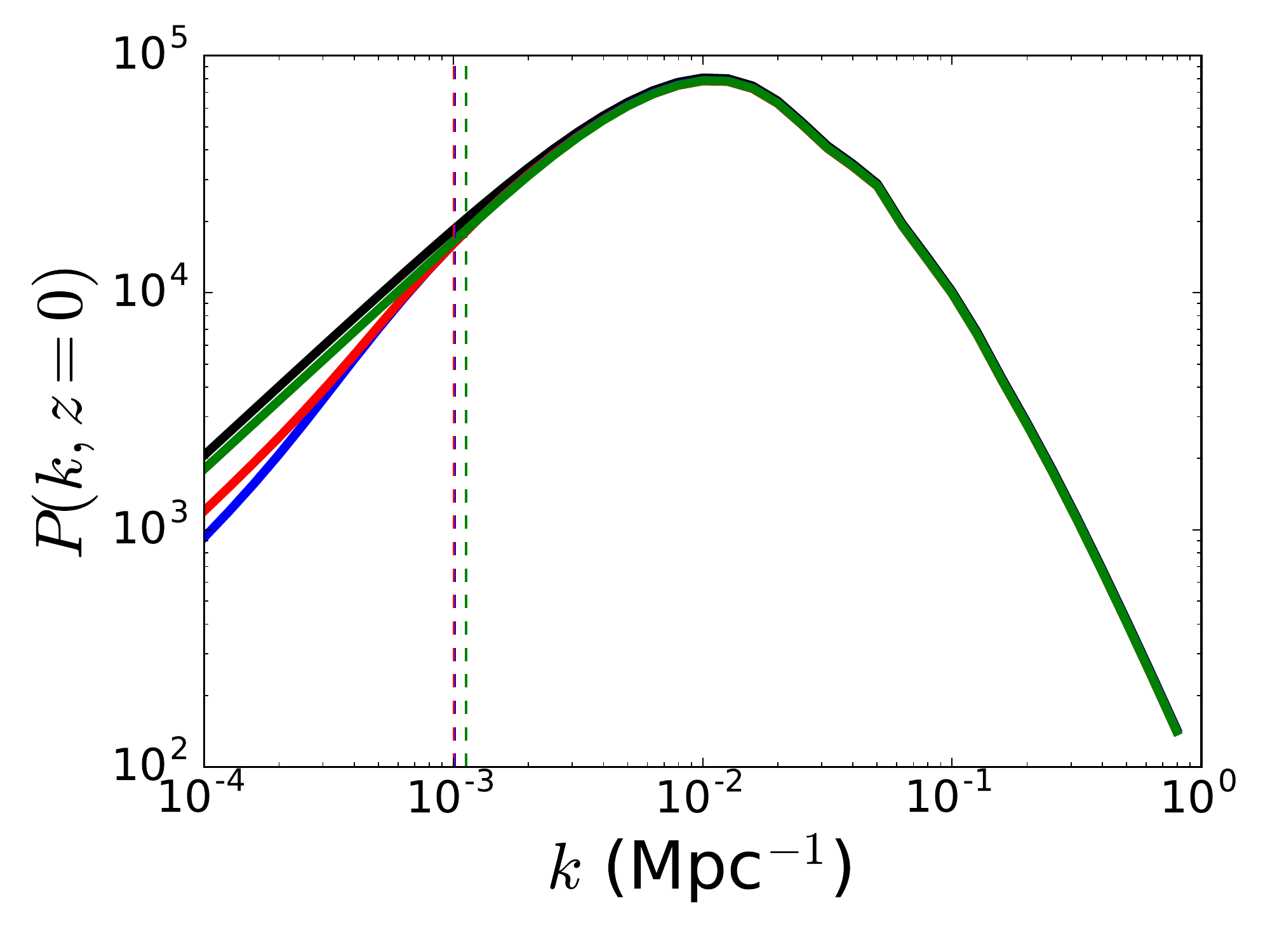}
\includegraphics[width=0.3\textwidth]{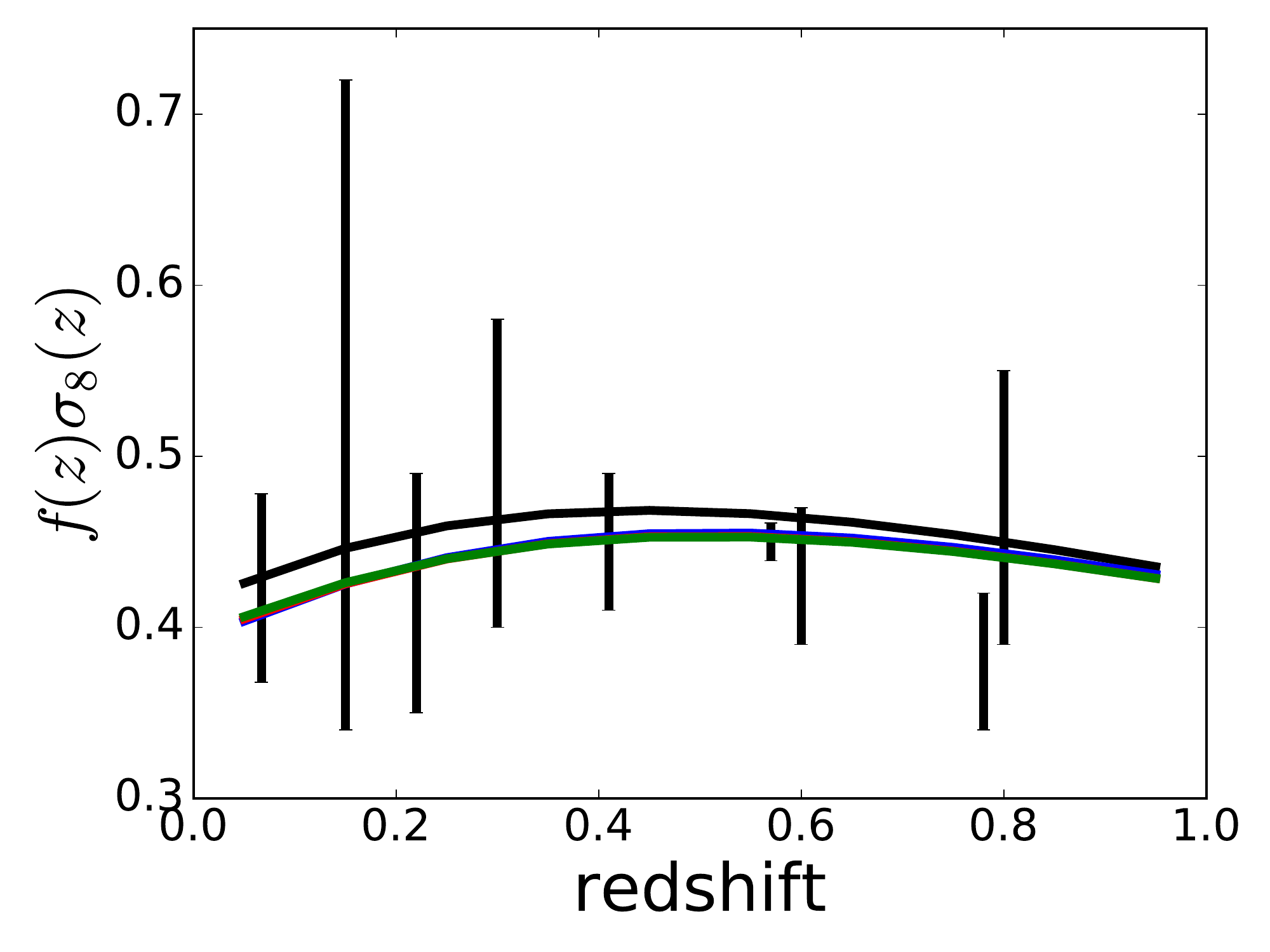}
\end{center}
\caption{CMB temperature power spectra, matter power spectra, and structure growth for $\Lambda$CDM and MG best fit models. From top to bottom, we show these quantities when we fit only the CMB, CMB+BAO, CMB+RSD, CMB+PK, and CMB+BAO+RSD+PK respectively. Vertical dashed lines correspond to the braiding scale $k_{\textrm{B}}(z=0)$ for each particular model and dataset combination. \label{fig:clpkfs8}}
\end{figure}

The best-fit models that correspond to the  Log Likelihood values of Tab.~\ref{tab:loglike} are shown in Figure \ref{fig:clpkfs8} where we present the cases in which we fit only CMB data, CMB and one of the Large Scale Structure datasets (BAO, RSD, or PK), and the full combination CMB+BAO+RSD+PK. We show how the observables fitted in our analysis are  reproduced by the MG models described in Section \ref{sec:method} where we fix the coefficient $c_\textrm{K}$ to 0 (blue lines), 1 (red lines), and 10 (green lines). We also include for direct comparison the best-fit model in a $\Lambda$CDM cosmology with standard gravity (black lines). We find that MG has more flexibility to fit the observables, but, given current error bars, the difference cannot be regarded   as more than a mere statistical fluctuation. It remains to be checked with future data whether the significance of these differences increases or is still consistent with noise.

Given the above apparent contradiction, we  investigate whether the posterior  confidence intervals  might be 
driven by strong correlations between the parameters varied in our MCMC chains. Strong correlations are known to introduce  prior volume effects in multidimensional analysis when the posterior distribution is non-Gaussian.

\begin{figure}
\begin{center}
\includegraphics[width=0.9\textwidth]{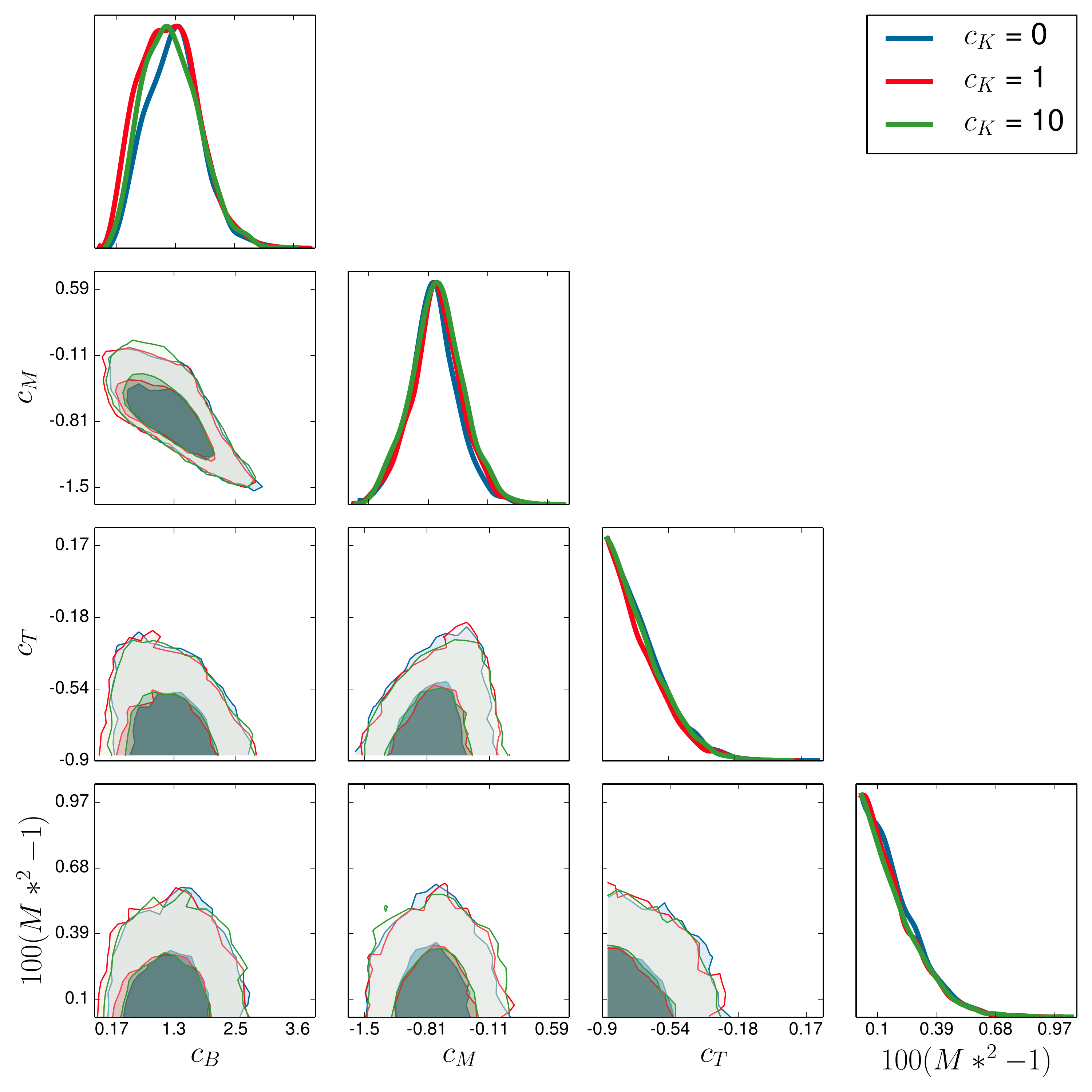}
\caption{Constraints on $c_\textrm{B}$, $c_\textrm{M}$, $c_\textrm{T}$ and $M^2_*$ from the combination of CMB+BAO+RSD+PK datasets.\label{fig:alphas}}
\end{center}
\end{figure}

Interestingly we find no significant correlations between the $\Lambda$CDM and the MG sectors of our parameter space, which greatly simplifies this analysis. We will therefore focus on possible correlations between the parameters of the MG sector. In figure~\ref{fig:alphas} we show the 68\% (light shade) and 95\% confidence level contours (dark shade) between pairs of MG parameters when the full CMB+BAO+RSD+PK dataset combination is used. 
This figure shows that, especially  in the $c_\textrm{B}$ case, it is the long non-Gaussian tail at high values that drives the effect.
The constraints on the MG parameters are nevertheless found to be uncorrelated, at least to the extent driven by the precision of the datasets used here.

The issue of whether the data favours a MG model is an issue of model selection or model comparison  rather than parameter estimation.
Within the Bayesian framework, which is the one adopted here, model comparison is done by considering the  model-averaged likelihood, referred to here as the Bayesian Evidence, $E$. Under the assumption of equal a priori model probabilities, the ratio of Evidence values for two models, given the same data, quantifies the relative odds of these models being the correct description of the observations.
Thus the key model comparison quantity is the Bayes factor, which is the ratio of the Evidence values for  the two  models. In general, the Evidence is the result of a multi-dimensional integral over the model parameters, the evaluation of which can be computationally expensive. However here the two models are {\it nested}: the simpler model (the $\Lambda$CDM+GR one) is a specific case of the  MG models  when the  four coefficients $c_{\textrm{B,M,T,K}}$ are zero. In this case it is possible to  perform rigorous model comparison between nested models without the need for evaluating numerically-intensive multi-dimensional integrals using the Savage-Dickey density ratio (SDDR; \cite{Dickey1971, VerdinelliWasserman95}).

\begin{table}
\begin{center}
\begin{tabular}{lcc}
   Dataset combination  & Evidence ratio $\ln \left( E_{\Lambda CDM}/E_{\textrm{H}} \right)$ & interpretation \\
\hline
CMB            & 1.48& substantial \\
CMB+BAO        &-0.21& not significant \\
CMB+RSD        & 0.64& not significant \\
CMB+PK         & 1.13& substantial \\
CMB+BAO+RSD+PK & 0.09& not significant \\
\hline
\end{tabular}
\caption{The Savage-Dickey Density Ratio for the  $\Lambda$CDM + GR  model  with respect to the Modified Gravity models studied here. We have considered the case $c_{\rm K}=0$. \label{tab:evidence}}
\end{center}
\end{table}

In Table~\ref{tab:evidence} we report the Bayes factor of   the $\Lambda$CDM  to  MG models  computed following \cite{Verde2013}; we use a  slightly modified version of the Jeffrey's scale to interpret the evidence ratios. The Bayes factor favours  the simpler, $\Lambda$CDM, model in the cases CMB and CMB+PK with odds $\sim 3:1$ (``substantial" evidence).

As with all Bayesian methods, the Bayes factor between the two models depends on the prior on the model parameters. The comparison between the Bayesian Evidence ratio and the Log Likelihood analysis presented above, which is prior-independent,  serves to  quantify a possible prior-dependence of the Bayes factor.

Next we  briefly discuss  the implications of our bounds on the $c_i$ parameters on the existence of a (possible) new scale associated to gravity (eq. \ref{braidingscale}).
 While formally one can compute a value for $k_B(z=0)$ given a set of parameters $c_i$, the interpretation of confidence levels on this parameter requires some caution.  As shown in Table~\ref{tab:lcdm},  the  $\Lambda$CDM limit (which is recovered exactly by setting all our parameters $c_i=0$) has an indeterminate $k_\textrm{B}$. In our analysis  the $\Lambda$CDM model offers a very good fit to the data. For the collection of models in parameter space close to   $\Lambda$CDM,  the $k_\textrm{B}$  distribution is unconstrained. When the $\Lambda$CDM model is not disfavoured by the data  the braiding scale is not a good  (derived) parameter to use  since small variations of our $c_i$ can produce huge changes on this scale.  For this reason we  do not dwell much on this quantity here. The braiding scale can still be a good quantity to use when investigating models  beyond the simple $\Lambda$CDM model: 
  it is meaningful  to evaluate it only in a MG scenario.

\begin{figure}
\begin{center}
\includegraphics[width=0.9\textwidth]{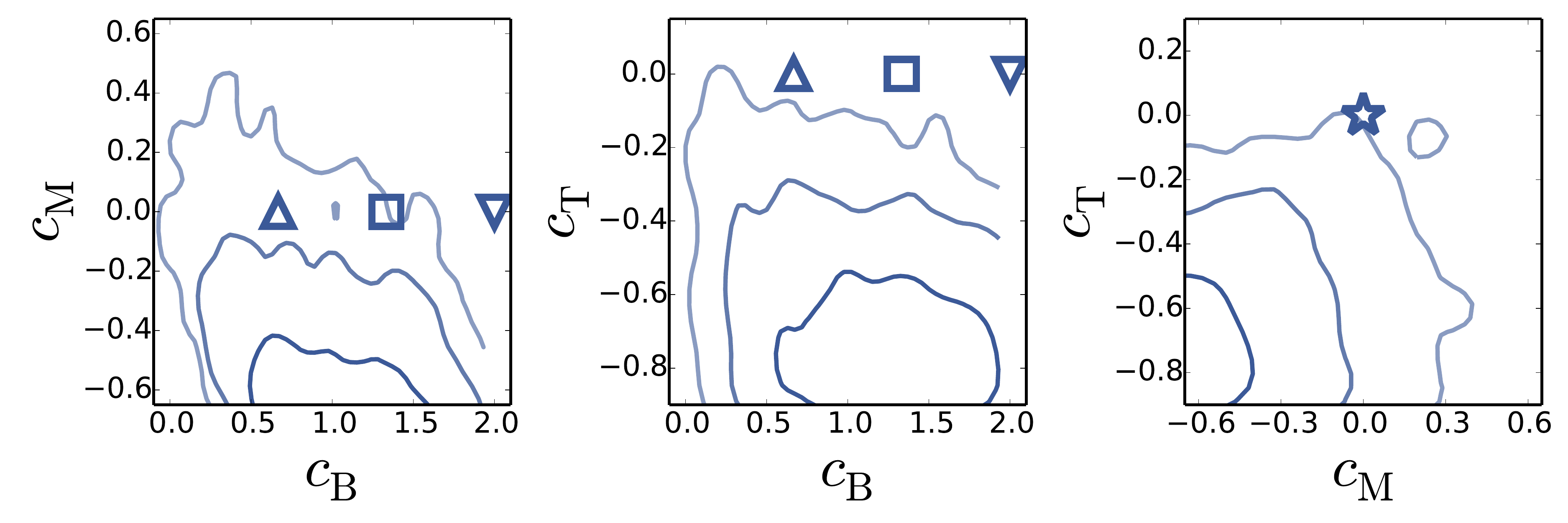}
\caption{Correlations between the coefficients $c_\textrm{B}$, $c_\textrm{M}$, and $c_\textrm{T}$. Contours are shown for 68.3\%, 95.4\%, and 99.7\% CL.  The symbols correspond to different values for the parameters of the model of Ref~\cite{Deffayet:2010qz,Pujolas:2011he} as discussed in the text. The three points in the left two panels correspond to (from left to right) $n=1/3$, $n=2/3$, $n=1$. In the right panel the three models overlap. Clearly  values $n\gtrsim 2/3$ are  disfavoured.  \label{FIG:correlations}}
\end{center}
\end{figure}

At this point it is useful to give a prescription on how to interpret our results for ``real'' theories. Indeed, strictly speaking the constraints we obtain are only valid for sub-classes of Horndeski that have the time evolution of their $\alpha$'s proportional to the time evolution of the DE/MG density on a $\Lambda$CDM background. In the literature, it is possible to find many models that are sub-classes of the Horndeski lagrangian, but in general each of them has a different dynamics. We claim that our results can be applied to more general frameworks. The reason for this is that most of the DE/MG models are constructed in such a way that they explain the late-time acceleration of the universe, with standard evolution at early-times (radiation/matter dominated eras) and modifications at late-times. Then, one can have in a reasonable amount of time (without the need of running MCMC chains) an idea of what is the allowed/not allowed region in the parameter space of a given theory. As a toy model we will make use of a minimal generalization of the imperfect fluid introduced in \cite{Deffayet:2010qz,Pujolas:2011he}. In this model we consider $K=-X^n$, $G_3=\mu X^m$, $G_4=M^2_\textrm{Pl}/2$ and $G_5=0$.\footnote{These functions are the usual Horndeski functions appeared first in \cite{Deffayet:2009mn}. The exact notation we used in order to define them can be found in Sec.~2, precisely Eq.~(2.1), of \cite{BS}. The model parameters therefore are $n$ and $m$.} This is a shift-symmetric model that has an attractor for the background evolution. As usual, on this attractor the equations can be integrated analytically, which is an advantage we use in order to illustrate some results. The steps one has to follow are:
\begin{enumerate}
 \item Solve the background evolution (i.e.~$H(t)$) and check that it is close to $\Lambda$CDM. Within our example it is possible to demonstrate that in the infinite future, where $\Omega_{DE}=1$, the equation of state for the scalar field is $w_{DE}=-1$;
 \item Use the definition of the alphas in terms of the Horndeski functions (in \cite{BS}) in order to express them as a function of background quantities. In our case we get: $\alpha_\mathrm{K}=6n(1-2n+2m)\Omega_{DE}$, $\alpha_\mathrm{B}=2n\Omega_{DE}$ and $\alpha_\mathrm{M}=\alpha_\mathrm{T}=0$;
 \item Check that the evolution of these  $\alpha$ functions is indeed $\propto\Omega_{DE}$ (or in which regime this applies). In our case this condition is satisfied exactly;\footnote{It is important to note that this condition can be satisfied approximately even in more general frameworks as the best fit models of the covariant galileons \cite{Barreira:2014jha,Bellini:2015oua}.}
 \item Invert the relations between our $c_i$ and the parameters of the theory (usually this step has to be done numerically). In our  the result is: $n=c_\textrm{B}/2$ and $m=(c_\textrm{K}/c_\textrm{B}+c_\textrm{B}-1)/2$;
 \item  Compare with constraints in  Fig.~\ref{FIG:correlations} to define the  allowed parameter region of the original theory.
\end{enumerate}

In Fig.~\ref{FIG:correlations} we illustrate this last step in the procedure. The symbols correspond to different values for the parameters of the toy model considered here. Values of $n\sim1/3$ or $2/3$ are allowed but  values $n \ge 3/4 $ are excluded. This is interesting as  $n=1$, which is the standard kinetic term  and standard case proposed in the original reference, is ruled out.

\begin{table}
\begin{center}
\begin{tabular}{lccc}
                         & $c_\textrm{B}$         & $c_\textrm{M}$         & $c_\textrm{T}$         \\
\hline
CMB+BAO+RSD+PK, $c_\textrm{K}=0$ & $-0.01<c_\textrm{B}<+3.06$ & $-1.65<c_\textrm{M}<+0.22$ & $-0.90<c_\textrm{T}<-0.13$ \\
CMB+BAO+RSD+PK, $c_\textrm{K}=1$ & $-0.12<c_\textrm{B}<+2.91$ & $-1.61<c_\textrm{M}<+0.27$ & $-0.90<c_\textrm{T}<-0.14$ \\
CMB+BAO+RSD+PK, $c_\textrm{K}=10$ & $+0.00<c_\textrm{B}<+2.86$ & $-1.59<c_\textrm{M}<+0.30$ & $-0.90<c_\textrm{T}<-0.15$ \\
\hline
\end{tabular}
\caption{Constraints on the coefficients $c_\textrm{B}$, $c_\textrm{M}$, and $c_\textrm{T}$ from different cosmological dataset combinations and for different values of $c_\textrm{K}$. Quoted limits are 99.73\% CL. A hard prior on $c_\textrm{T}>-0.9$ is applied.\label{tab:alphas3sigma}}
\end{center}
\end{table}

\section{Discussion and conclusions}
\label{sec:conclu}

We have considered a  general version of Horndeski theories of gravity,  where modifications to general relativity on cosmological scales are well described by an additional scalar degree of freedom with at most second-order derivatives in the equations of motion and where  the weak equivalence principle is satisfied. This description encompasses many of the classical DE/MG models studied to explain the late-time cosmic acceleration. We chose a parametrization in which the time evolution of the four arbitrary functions in the Horndeski Lagrangian is given by the time evolution of $\Omega_{DE}$. The $\Lambda$CDM limit is recovered when all the extra parameters are zero.

 We have performed a  global fit for the parameters of the model (including the cosmological parameters), considering the state-of-the art cosmological data: Cosmic Microwave Background data from  the Planck mission and the matter power spectrum, redshift-space distortions and Baryon Acoustic Oscillation data from galaxy surveys of  Large Scale Structure.

We used the standard Bayesian approach to parameter inference and obtained posterior confidence intervals on the model's parameters.
The  main result of this work is that  we find no significant  statistical evidence for deviations from Einstein's gravity in the current data.  While formally  for some data-sets combinations the $\Lambda$CDM+GR limit is (just) outside the 95\% posterior  confidence interval, the improvement in the fit at the expense of adding extra parameters, quantifies in terms of difference of log likelihood is not significant. 
The Bayes factor also confirms  this by not favouring the more complex model  (Horndeski gravity) over the simpler  ($\Lambda$CDM+GR) one. These findings are robust to to choice of  specific data-set used and  deviations from  the $\Lambda$CDM+GR limit are  mostly  driven by the low multipoles of the Cosmic Microwave Background anisotropies.

These results are the best constraints so far on this parametrization of the Horndeski Lagrangian, in which $\alpha_i \propto \Omega_{DE}$, which limit  any possible deviations from $\Lambda$CDM+GR.

It is interesting to note that in constraining this class of models beyond $\Lambda$CDM, the measurement of large-scale  Cosmic Microwave Background polarization plays an important role. In fact we find that the low $\ell$ polarization data from the Planck satellite are less constraining than the WMAP ones. Forthcoming improvements on large-scale polarization measurements might help improve our constraints.

While strictly speaking the constraints we obtain are only valid for sub-classes of Horndeski considered,  our results
can be applied to a more general framework.  In fact most of the DE/MG models reported in the literature are
constructed in such a way to explain the late-time acceleration of the universe, with standard
evolution at early-times (radiation/matter dominated eras) and modifications at late-times, making therefore possible to ``map" them  to our class of model by interpreting our parameters as {\it effective} parameters. This mapping is of course model-dependent and might not always be analytic, but this procedure enables one to  rule in or out specific models or parameter ranges for models without having to perform any data analysis or expensive exploration of multidimensional parameter-space and  inference.  We  have illustrated this procedure for a specific MG model.
\acknowledgments
We are very grateful to Miguel Zumalac\'arregui and the \textsc{HiClass} \cite{hi-class:2015} collaboration for letting us use this code, essential for the analysis of this paper. We also want to thank Janina Renk and Miguel Zumalac\'arregui for valuable comments and criticisms on the manuscript. This work was supported in part by the Radcliffe Institute for Advanced Study at Harvard University. RJ  and LV thank the Royal Society  for financial support and  the ICIC at Imperial College for hospitality while this work was being completed. LV, EB and AJC are supported by the European Research Council under the European Community's Seventh Framework Programme FP7-IDEAS-Phys.LSS 240117. 
Funding for this work was partially provided by the Spanish MINECO under projects AYA2014-58747-P and MDM-2014-0369 of ICCUB (Unidad de Excelencia 'Mar{\'\i}a de Maeztu'). Based on observations obtained with Planck (\url{http://www.esa.int/Planck}), an ESA science mission with instruments and contributions directly funded by ESA Member States, NASA, and Canada.

Funding for SDSS-III has been provided by the Alfred P. Sloan Foundation, the Participating Institutions, the National Science Foundation, and the U.S. Department of Energy Office of Science. The SDSS-III web site is http://www.sdss3.org/.
SDSS-III is managed by the Astrophysical Research Consortium for the Participating Institutions of the SDSS-III Collaboration including the University of Arizona, the Brazilian Participation Group, Brookhaven National Laboratory, Carnegie Mellon University, University of Florida, the French Participation Group, the German Participation Group, Harvard University, the Instituto de Astrofisica de Canarias, the Michigan State/Notre Dame/JINA Participation Group, Johns Hopkins University, Lawrence Berkeley National Laboratory, Max Planck Institute for Astrophysics, Max Planck Institute for Extraterrestrial Physics, New Mexico State University, New York University, Ohio State University, Pennsylvania State University, University of Portsmouth, Princeton University, the Spanish Participation Group, University of Tokyo, University of Utah, Vanderbilt University, University of Virginia, University of Washington, and Yale University.

\bibliographystyle{utcaps}
\bibliography{nus}
\section*{Appendix A: Constraining power of different CMB datasets}
\label{sec:appendix}

In this Appendix we justify our default choice for the CMB dataset combination. In Table~\ref{tab:cmbcomp} we show how our constraints change when we choose a different CMB dataset combination. Although the Planck 2015 case seems inconsistent with the others (especially for $c_{\textrm B}$), it also forces the optical depth to reionization $\tau_{\textrm{reio}}$ to be less than 0.04, which we know to be inconsistent with other cosmological observations such as e.g., the Gunn-Peterson effect (see e.g., \cite{Caruana}) or WMAP data themselves. We conclude that the most constraining CMB dataset while consistent with  other cosmological observations is the one used in our default case.

\begin{table}
\begin{center}
\begin{tabular}{lccc}
                         & $c_\textrm{B}$         & $c_\textrm{M}$         & $c_\textrm{T}$         \\
\hline
Planck13+WP+BAO+RSD+PK,        $c_\textrm{K}=0$ & $+0.10<c_\textrm{B}<+2.86$ & $-1.57<c_\textrm{M}<+0.34$ & $-0.90<c_\textrm{T}<-0.21$ \\
Planck15+lowTEB+BAO+RSD+PK,    $c_\textrm{K}=0$ & $+0.35<c_\textrm{B}<+2.86$ & $-1.55<c_\textrm{M}<+0.01$ & $-0.90<c_\textrm{T}<-0.25$ \\
Planck15+tauprior+BAO+RSD+PK,  $c_\textrm{K}=0$ & $+0.02<c_\textrm{B}<+2.10$ & $-1.01<c_\textrm{M}<+0.43$ & $-0.90<c_\textrm{T}<-0.03$ \\
Planck15+lowT+WP+BAO+RSD+PK,   $c_\textrm{K}=0$ & $+0.24<c_\textrm{B}<+2.32$ & $-1.36<c_\textrm{M}<-0.13$ & $-0.90<c_\textrm{T}<-0.39$ \\
\hline
Planck13+WP+BAO+RSD+PK,        $c_\textrm{K}=1$ & $+0.02<c_\textrm{B}<+2.75$ & $-1.52<c_\textrm{M}<+0.29$ & $-0.90<c_\textrm{T}<-0.21$ \\
Planck15+lowTEB+BAO+RSD+PK,    $c_\textrm{K}=1$ & $+0.34<c_\textrm{B}<+2.90$ & $-1.57<c_\textrm{M}<+0.04$ & $-0.90<c_\textrm{T}<-0.25$ \\
Planck15+tauprior+BAO+RSD+PK,  $c_\textrm{K}=1$ & $-0.03<c_\textrm{B}<+2.01$ & $-0.91<c_\textrm{M}<+0.37$ & $-0.90<c_\textrm{T}<-0.01$ \\
Planck15+lowT+WP+BAO+RSD+PK,   $c_\textrm{K}=1$ & $+0.10<c_\textrm{B}<+2.29$ & $-1.35<c_\textrm{M}<-0.08$ & $-0.90<c_\textrm{T}<-0.41$ \\
\hline
Planck13+WP+BAO+RSD+PK,        $c_\textrm{K}=10$ & $+0.13<c_\textrm{B}<+2.75$ & $-1.51<c_\textrm{M}<+0.35$ & $-0.90<c_\textrm{T}<-0.22$ \\
Planck15+lowTEB+BAO+RSD+PK,    $c_\textrm{K}=10$ & $+0.35<c_\textrm{B}<+2.94$ & $-1.58<c_\textrm{M}<+0.01$ & $-0.90<c_\textrm{T}<-0.24$ \\
Planck15+tauprior+BAO+RSD+PK,  $c_\textrm{K}=10$ & $-0.02<c_\textrm{B}<+1.97$ & $-0.89<c_\textrm{M}<+0.36$ & $-0.90<c_\textrm{T}<-0.02$ \\
Planck15+lowT+WP+BAO+RSD+PK,   $c_\textrm{K}=10$ & $+0.19<c_\textrm{B}<+2.30$ & $-1.36<c_\textrm{M}<-0.06$ & $-0.90<c_\textrm{T}<-0.41$ \\
\hline
\end{tabular}
\caption{Constraints on the coefficients $c_\textrm{B}$, $c_\textrm{M}$, and $c_\textrm{T}$ from different Cosmic Microwave Background datasets in combination with BAO+RSD+PK and for different values of $c_\textrm{K}$. Quoted limits are 95\% CL. A hard prior on $c_\textrm{T}>-0.9$ is imposed.\label{tab:cmbcomp}}
\end{center}
\end{table}

\end{document}